\documentclass{JHEP3}
\usepackage{amsmath,amssymb}
\usepackage[dvips]{graphicx}
\usepackage{epsfig}
\usepackage{psfrag}

\title{Dissolving D0-brane into D2-brane with background B-field} 
\author{Yoshinao Sato\\
Institute of Physics, University of Tokyo\\
Komaba, Meguro-ku, Tokyo 153-8902, Japan
\\E-mail: \email{ysato@hep1.c.u-tokyo.ac.jp}}

\preprint{UT-Komaba/05-4}

\abstract{ D0-branes on a D2-brane with a constant background B-field are
unstable due to the presence of a tachyonic mode and expected to
dissolve into the D2-brane to formulate a constant D0-charge density. In this paper we study such a dissolution process in terms of a noncommutative gauge theory. Our results show that the localized D0-brane spreads out over all of space on the D2-brane as the tachyon rolls down into a stable vacuum. D0-branes on a D2-brane can be described as unstable solitons in a noncommutative gauge theory in 2+1 dimensions in the Seiberg-Witten limit. In contrast to the case of annihilation of a non-BPS D-brane, we are free from difficulty of disappearance of DOF, since there exist open strings after the tachyon condensation. We solve an equation of motion of the gauge field numerically, and our results show that the localized soliton smears over all of noncommutative space. In addition, we evaluate distributions of D-brane charge, F-string charge, and energy density via formulas derived in Matrix theory. Our results show that the initial singularities of D0-charge and energy density are resolved by turning on the tachyon, and they disperse over the whole space on the D2-brane during the tachyon condensation process. }

\begin{document}
\section{Introduction}
Recently the study of a decay process of non-BPS D-branes have made considerable developments and revealed a new aspect of the relation between open and closed degrees of freedom (DOF) \cite{0203211,0203265,0204143,0205085,0205098,0209034,0303139,0306137,0410103}. In oder to understand a role of open and closed string DOF in D-brane dynamics more precisely, it is meaningful to study a time-dependent process of a D-brane system in which an effective theory description is valid during the whole time, and disappearance of open string DOF does not happen. In this paper we study a tachyon condensation in a D0/D2-brane system, or a process in which a D0-brane dissolves into a D2-brane with a constant background B-field, in terms of a noncommutative gauge theory.

Decay of a non-BPS D-brane was studied in
terms of tachyonic effective field theories of several types in addition to
a boundary CFT analysis. These effective theories, however, do not
necessarily have a rigorous justification from the viewpoint of string theories. 
We would not naively expect a low energy effective theory since the tachyon has mass squared of the order $1/\alpha'$. In addition, there is no S-matrix to check the validity of an effective theory at a stable vacuum without any brane. On the other hand, D0-branes on a D2-brane with a constant background B-field \cite{9704006,0105079} can be
described as noncommutative gauge solitons in 2+1 dimensions \cite{0007204,0009142,0010090} in the Seiberg-Witten limit \cite{9908142}. Such D0-branes are unstable because of a tachyonic mode\footnote{Stability of D0-D$p$ systems was studied in \cite{0105079}. Scattering of D0-branes was also investigated \cite{0112050}. Different kinds of D$p$-D$q$-brane systems and noncommutative solitons have been studies in \cite{0103196,0106213,0310267,0504025}.} and is expected to dissolve into the D2-brane. Mass squared of the tachyon is not of order $1/\alpha'$ and it survives in the Seiberg-Witten limit. Since we have a D2-brane with a dissolved D0-brane as remnant after the tachyon condensation, we do not suffer from the difficulty of death of open strings. In the tachyon condensation process, 0-0 and 0-2 open string modes does not disappear, but metamorphose
to 2-2 open string modes. 

In a D0/D2-brane system there exist an infinite number of massless and massive modes which survive in the Seiberg-Witten limit in addition to a tachyon mode\footnote{Since we take the Seiberg-Witten limit, a stringy massive tower of energy scale $1/\alpha'$ is eliminated. We can say that we treat infinitely many light modes with a fine splitting structure due to the presence of magnetic flux.}. 
Then we cannot concentrate on dynamics of the tachyon, but need to consider all other modes in order to study a tachyon condensation process in this system. However we can truncate the number of DOF in a way which has a physical interpretation in contrast to the level truncation in a string field theory. A D0/D2-brane system can be described by using a matrix-valued configuration of infinitely many D0-branes. As is seen easily
in such a viewpoint, we have infinitely many modes associated to an infinite number of matrix components. Since it is difficult to deal with infinitely many modes, we truncate the matrix to a finite size. Such matrix truncation has a physical
interpretation that the noncommutative space on the D2-brane is ``cut off''. Taking the limit of matrix size to infinity, the whole worldvolume is recovered. In other words
we treat only a finite number of modes to get
results which are reliable within a corresponding distance from the initial position of the D0-brane on the D2-brane.

In this paper we restrict ourselves to the rotationally symmetric case which is natural to understand the physics of this system. We numerically solve an equation of motion with a Gauss law constraint for the gauge field to study the time evolution of the unstable soliton. Our results show that the localized soliton disperses or smears over the whole space on the D2-brane. At earlier time the tachyoin mode and lower number state modes of the gauge field, which we define in section \ref{section:2} (see also footnote \ref{footnote:1} in section \ref{section:3}), roll down to the potential bottom. As time goes higher number state modes sequentially approach the stable vacuum, and then sudden disturbance happens. Considering that higher/lower number state modes correspond to more dispersed/localized configurations of the gauge field, this is a process in which a configuration of the gauge field disperses in the noncommutative space. When the dissolved D0-brane expands beyond the ``cut off'' domain, a phenomenon to say ``reflection'' to inside happens. In the limit of infinite matrix size, we expect that no ``reflection'' or sudden disturbance happens and the configuration of the gauge field disperses forever. 

Furthermore we evaluate supergravity charge distribution via formulas derived in Matrix theory \cite{9712185,9812239,0012218,0103124} in order to understand this process geometrically. In a noncommutative gauge theory there do not exist local and gauge invariant observables since we cannot distinguish between gauge transformations and translations in noncommutative directions. We can, however, extract gauge invariant observables from a matrix-valued configuration by using these formulas \cite{0008075,0104036}. Our results of evaluation show that the initial singularities of D0-charge and energy density are resolved by turning of the tachyon and they disperse over all of space on 
 the D2-brane as the system rolls down to the stable vacuum. Contributions of the tachyon to D0-charge and energy density distributions include a negative infinity at the D0-brane position, and this negative divergence is expected to cancel out the initial delta function singularity. All other modes cause polynomials with Gaussian factors of the distance from the initial position of the D0-brane, and high number state modes correspond to charge distributions at far distance. Thus the region in which D0-charge and energy density have non-zero value always enlarges as the D0-brane dissolves. A D2-charge distribution does not change for any configuration and F-charge density formulates a non-zero distribution with no divergence. 
We can verify that during the whole process the total charges and energy are conserved. To evaluate these charges we need to compute a symmetrized trace of operators which include exponential of an infinite size matrix. We provide prescriptions to calculate such formulas in expansion with respect to small fluctuations around the soliton. 

Our study is related particularly to papers cited below. Noncommutative gauge theories describing a D0/D2-brane system and a D1/D3-brane system were studied in detail in \cite{0007204,0008204,0009142,0010090,0011099}. In a T-dual language, dissolution of D0-brane corresponds to reconnection or recombination of intersecting D1-strings, which was studied in terms of the $SU(2)$ Yang-Mills theory in 1+1 dimensions in \cite{0303204,0304237}. Charge distributions of various noncommutative solitons were calculated in \cite{0105311} via the same formulas used in this paper. Tachyon condensation in D0/D4-brane system is studied in \cite{0502158}. Rolling tachyon in other noncommutative field theories was studied in \cite{0301119,0501016}.
 
In section \ref{section:2}, we briefly review a noncommutative gauge theory on a D2-brane with a background B-field, and specify the tachyon condensation process to be studied.  Time evolution of tachyon rolling and distribution of supergravity charge density are studied in section \ref{section:3}. We also provide prescriptions to calculate charge density formulas in section \ref{section:3}. Finally we conclude our study in this paper and discuss future directions in section \ref{section:4}. In appendix \ref{append:a} we review spectrum matching between a noncommutative gauge theory and a D0/D2-brane system. Some of equations which are used to evaluate charge density are proven in appendix \ref{append:b}, and detailed calculations of charge density are presented in appendix \ref{append:c}. In appendix \ref{append:d} we see a relation between a D0/D2-brane and intersecting D1-strings \cite{0303204,0304171,0304237} in the Seiberg-Witten limit.

\section{Noncommutative gauge field theory}\label{section:2}
In this section we briefly review a noncommutative gauge theory \cite{9701078,0011095,0102076,0106048,0109162} and prepare our setup\footnote{Concerning noncommutative instantons and solitons see also \cite{9802068,0003160,0005031,0005204,0007043,0007217,0007226,0010060}.}. Open strings on a D2-brane with background B-field are described in the Seiberg-Witten limit by a $U(1)$ noncommutative gauge theory in 2+1 dimensions \cite{0007204,0008204,0009142,0010090,0011099}. We denote noncommutative coordinates as $\hat{x}^i \ (i=1,2)$ and a noncommutative parameter as $\theta^{ij}=\theta \epsilon^{ij}$ which satisfy
\begin{equation*} 
   [\hat{x}^i,\hat{x}^j]=i\theta^{ij}
\end{equation*}
where $\epsilon$ is the antisymmetric tensor with $\epsilon^{12}=1$ and $\theta >0$. We introduce covariant coordinates
\begin{equation*}
  X^i=\hat{x}^i+\theta^{ij}A_j(\hat{x})
\end{equation*}
and connection operators
\begin{equation*}
   C=\frac{1}{\sqrt{2\theta}}(X^1-iX^2) \qquad \bar{C}=\frac{1}{\sqrt{2\theta}}(X^1+iX^2).
\end{equation*}
It is convenient to introduce creation and annihilation operators
\begin{equation*}
  a^\dagger=\frac{1}{\sqrt{2\theta}}(\hat{x}^1-i\hat{x}^2) \qquad a=\frac{1}{\sqrt{2\theta}}(\hat{x}^1+i\hat{x}^2). 
\end{equation*}
Here $a^\dagger$ and $C$ are defined as dimensionless operators. Functions of noncommutative coordinates can be identified with operators, called Weyl symbols, acting on Fock space $\mathcal{H}$ generated by the creation operator from the vacuum:
\begin{align*} 
   \mathcal{H}&=\mathbb{C} \{|n\rangle \ | \ n= 0,1,2,\cdots\} \\
    |n\rangle&=\frac{(a^\dagger)^n}{\sqrt{n!}}|0\rangle \quad a^\dagger a|n\rangle=n |n\rangle  \quad a|0\rangle =0.
\end{align*} 
Thus the physics of this system is described by operators acting on $\mathcal{H}$. The function of which Weyl symbol becomes $|n\rangle\langle m|$ is computed as \cite{0102076} 
\begin{equation}
   f_{nm}(\hat{x}^1,\hat{x}^2)=(-1)^n 2\sqrt{\frac{n!}{m!}}e^{-\hat{r}^2}(2\hat{r}^2)^{\frac{m-n}{2}}e^{i\hat{\varphi}(m-n)}L_n^{m-n}(2\hat{r}^2), \label{eq:Laguerre}
\end{equation}
where $\hat{x}^1+i\hat{x}^2=\hat{r}e^{i\hat{\varphi}}$ and $L_n^k$ is the associated Laguerre polynomial, which is a $n$-th order polynomial. An action of U(1) noncommutative Yang-Mills theory in temporal gauge $A_0=0$ is written as
\footnote{At a glance $\theta^{ij}$ seems to lead only an infinite constant shift of the action. However one would have a mistake assuming $\mathrm{tr}[C,\bar{C}]=0$ for infinite size matrices. Actually $\mathrm{tr}[C,\bar{C}]$ should be considered as a topological quantity. Thus we cannot drop $\theta^{ij}$ in the action \cite{0106048}.}
\begin{equation}
\begin{split}
   S=&\frac{2\pi}{g_\mathrm{YM}^2\theta} \int \!\! dt \sqrt{|G|} \ \mathrm{Tr}  \left[ \frac{1}{2}(\partial_t X^i)^2-\frac{1}{4\theta^2}(i[X^i,X^j]+\theta^{ij})^2\right]\\
   =&\frac{2\pi}{g_\mathrm{YM}^2}\int \!\! dt \mathrm{Tr} \left[\partial_t \bar{C} \partial_t C-\frac{1}{2G\theta}([C,\bar{C}]+1)^2 \right]. \label{eq:action}
\end{split}
\end{equation}
In order to consider correspondence with a string theory, the open string metric $G$ defined as $G_{\mu\nu}=G\delta_{\mu\nu}$ is not set to 1. Extension of the gauge group to U($N$) by tensoring Fock space $N$ times corresponds to coincident N D2-branes in a string theory interpretation. The equation of motion for $C$ becomes
\begin{equation}
   \partial_t^2 C = \frac{1}{G\theta}[C,[C,\bar{C}]]. \label{eq:EOM}
\end{equation}
The temporal gauge imposes a Gauss law constraint written as
\begin{equation} 
   [\bar{C},\partial_t C]+[C,\partial_t\bar{C}]=0. \label{eq:Gauss}
\end{equation}
A solution $C=a^\dagger$ represents a stable vacuum with vanishing gauge field, or no D0-brane as seen below. 
A series of static solutions of \eqref{eq:EOM} labeled by a positive integer $m$, which can be interpreted as $m$ D0-branes, is given by the form
\begin{equation*} 
   C=U a^\dagger U^\dagger +\sum_{i=0}^{m-1} \mu_i |i \rangle\langle i|.
\end{equation*}
Here $\mu_i$ is a complex parameter relating to a position of D-branes, and $U, U^\dagger$ are shift operators defined as $U=\sum_{k=0}^{\infty}|k\rangle \langle k+m|, U^\dagger=\sum_{k=m}^{\infty} |k\rangle \langle k-m|$. Note that shift operators satisfy equations
\begin{equation*}
   UU^\dagger =1 - \sum_{k=0}^{m-1}|k\rangle\langle k|, \quad U^\dagger U=1.
\end{equation*}
These localized solutions have an energy
\begin{equation*} 
   E=\frac{m\pi}{g_\mathrm{YM}^2 G \theta}.
\end{equation*}
These solitons are unstable because of a tachyonic mode, and so
expected to decay into the stable vacuum to formulate smeared
solutions. Before we investigate such a process of tachyon condensation, we
calculate a mass spectrum around the unstable soliton and the potential of this system. In the remaining of this paper, we focus on the case of $m=1$, or a single D0-brane. 

In the number operator representation, the soliton solution is written as
\begin{align*}
   C_0&=\begin{pmatrix} \mu & 0 \\ 0 & a^\dagger \end{pmatrix}\\
   a^\dagger&=\sum_i \sqrt{i+1}|i+1 \rangle\langle i|. 
\end{align*}
We denote fluctuations around this solution, using submatrices $A, W, T^\dagger, D$ of which components correspond to respectively 0-0, 0-2, 2-0, 2-2 strings, as
\begin{gather*} 
   \delta C =\begin{pmatrix}A & W \\ T^\dagger & D  \end{pmatrix} \\
   W=(w_0 \ w_1 \ \cdots) ,\quad  T^\dagger=\begin{pmatrix} t_0 \\ t_1 \\ \vdots \end{pmatrix}, \quad D=\begin{pmatrix} d_{00} & d_{01} & \\ d_{10} & d_{11} & \cdots \\ & \vdots & \end{pmatrix}.
\end{gather*} 
Then the whole operator $C$ is given by
\begin{equation*}
   C=
   \begin{pmatrix}
     \mu+A & W \\ T^\dagger & a^\dagger+D 
   \end{pmatrix}.  
 \end{equation*}
Here the size of the submatrix $A$ is set to $m \times m$. Particularly in our case of $m=1$, $A$ is just a complex number, which shifts a position of the single D0-brane. As a consequence of the Gauss law constraint, all of theses fluctuations are not necessarily independent.

In this paper we restrict ourselves to the rotationally symmetric case. In what follows we find which DOF survives in this restriction and calculate the potential. For noncommutative coordinates, derivative operators are formally written as $\hat{\partial}_i=-i\theta_{ij}\hat{x}^j$ where $\theta_{ij}$ is the inverse of $\theta^{ij}$, namely $\theta_{ij}\theta^{jk}=\delta_i^k$. Thus rotation in the Fock space is generated by
\begin{equation*}
   \hat{x}^1\hat{\partial}_2-\hat{x}^2\hat{\partial}_1=-i\frac{\hat{x}^2}{\theta}=-2i\left(a^\dagger a +\frac{1}{2}\right),
\end{equation*}
so that
\begin{align*}
    \hat{O} &\quad \to \quad  R\hat{O}R^\dagger \\
    R&=\exp(-i\phi a^\dagger a),   
\end{align*}
denoting an operator in Fock space as $\hat{O}$. Notice that the number operator $a^\dagger a$ and an operator $\hat{x}^2=(\hat{x}^1)^2+(\hat{x}^2)^2$ obeys a relation $\frac{\hat{x}^2}{2\theta}=a^\dagger a +\frac{1}{2}$. We can see $\hat{x}^i$ is rotated by the angle $\phi$ under the rotation operator $R$ as
\begin{equation*}
   R\hat{x}^iR^\dagger=T^i_j \hat{x}^j  \qquad T^i_j=\begin{pmatrix}\cos \phi & \sin \phi \\ -\sin \phi & \cos \phi \end{pmatrix}
\end{equation*}
or equivalently $Ra^\dagger R^\dagger = e^{-i\phi} a^\dagger$. We can see that the unstable soliton $C=U a^\dagger U^\dagger +\mu |0\rangle\langle 0|$ is rotationally symmetric only when $\mu=0$ or the D0-brane is located at the original point $x^i=0$. Next we consider which component in $C=a^\dagger-i\theta A_z$ must be zero in the rotationally symmetric case. Assuming gauge field configurations to be rotationally symmetric, $A^i$ should be rotated as $RA_iR^\dagger=T_i^jA_j$, or equivalently $RA_zR^\dagger=e^{-i\phi}$. Hence $C$ must be rotated as
\begin{equation}   
   RCR^\dagger=e^{-i\phi}C. \label{eq:rotate}
\end{equation}
in the rotationally symmetric case. Decompose a Fock space operator as $\hat{O}=\sum_{ij} O_{i,j}|i \rangle\langle j|$, then the rotation operator acts on $\hat{O}$ as
\begin{equation*}
  R\hat{O}R^\dagger=\sum_l e^{-il\phi }\left(\sum_i O_{i+l,i}|i+l\rangle\langle i|\right).    
\end{equation*}
From this equation we can say that an operator of the form $\hat{O}_l=\sum_i O_{i+l,i}|i+l \rangle \langle i|$ is rotated to have a phase factor $e^{-il\phi}$ \footnote{ This result can be understood also from the equation \eqref{eq:Laguerre}. }. Thus \eqref{eq:rotate} requires that $\delta C_{ij}=0$ unless $i-j=1$ if we want to turn on the tachyon. Then the operator $C$ can be written as
\begin{equation}    
          C = \begin{pmatrix} 0   & 0     & 0            & 0 &       \\
                              t_0 & 0     & 0            & 0 & \cdots\\ 
                              0   & 1+d_0 & 0            & 0 &       \\
                              0   & 0     & \sqrt{2}+d_1 & 0 &       \\
                                  &\vdots &              &   & 
\end{pmatrix}. \label{eq:fluctuation}
\end{equation}
where we have defined $d_{i-1}=d_{i,i-1}$. Pay attention that we call $d_i$ high number state modes when $i$ is sufficiently large, and low number state modes when $i$ is sufficiently small, although this terminology is not strict. After all the potential in the rotationally symmetric case is given by
\begin{equation*}
    V=\frac{1}{2G\theta}\{(|t_0|^2-1)^2 + (|t_0|^2-|d_0+1|^2+1)^2 + \sum_{i=1}
    (|d_{i-1}+\sqrt{i}|^2-|d_i+\sqrt{i+1}|^2+1)^2 \}. 
\end{equation*}
It can be easily seen that the potential is positive definite and $C=a^\dagger$, or $t_0=1,d_i=\sqrt{i+2}-\sqrt{i+1}$, is the stable vacuum. In addition we can read off that mass squared of $t_0$ is\footnote{Considering $C=a^\dagger$ corresponds to $A_z(x)=0$, $d_i$ can be identified as massless modes of the gauge field \cite{0007204,0502158}. To diagonalize quadratic terms of $d_i$ in the potential with matrix truncation is complicated.}
\begin{equation*}
   m^2=-\frac{1}{G\theta}.
\end{equation*}
In this paper we analyze a process of the tachyon condensation from an unstable soliton to the stable vacuum 
\begin{alignat*}{4}
  & C=Ua^\dagger U^\dagger & \qquad \to \qquad       & C=a^\dagger& \\
  & t_0=0,d_i=0&                      & t_0=1,d_i=\sqrt{i+2}-\sqrt{i+1} &. \nonumber
\end{alignat*}
We can see that fluctuations around the potential top $t_0,d_0,d_1,\cdots$ metamorphose to fluctuations which correspond to $d_0,d_1,d_2,\cdots$ around the potential bottom. Thus mass spectrum around the solution $C_0=a^\dagger$ is same as the one around the localized solution $C_0=Ua^\dagger U^\dagger$ except the tachyon.

\EPSFIGURE[htb]{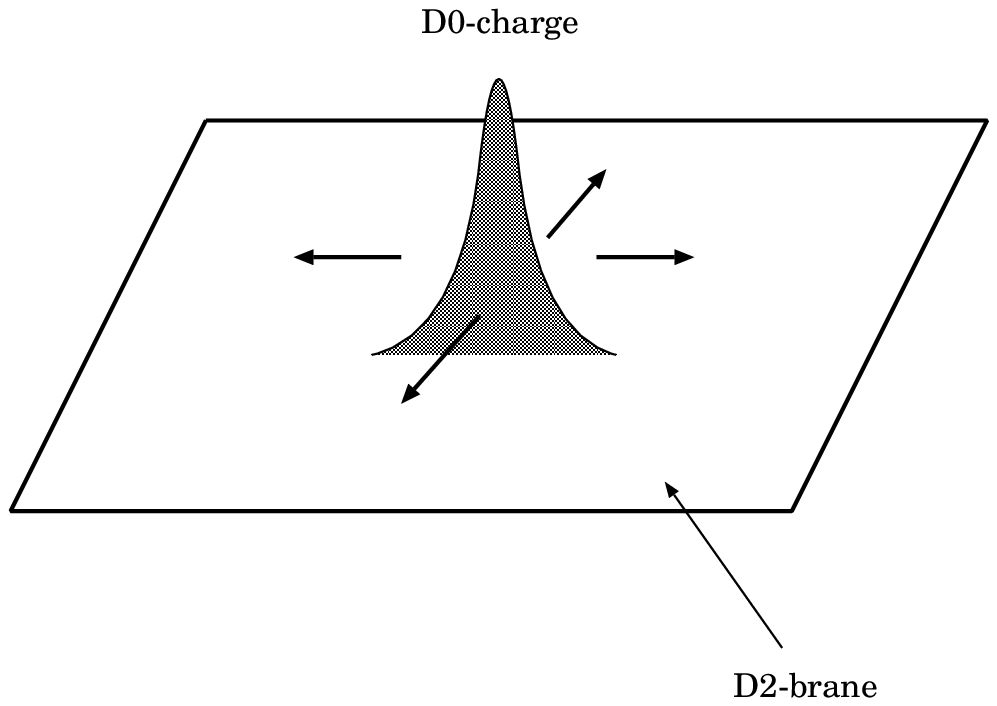}{Dissolving D0-brane}

In the remaining of this section we briefly review the Seiberg-Witten limit. The Seiberg-Witten limit is a zero slope limit $\alpha'\to 0$ keeping open string parameters $G,\theta$. When we denote a zero slope limit as $\alpha' \to \epsilon^{1/2}$ the Seiberg-Witten limit can be written as
\begin{gather}
   \alpha' \to \epsilon^{1/2}, \quad
   g \to \epsilon, \quad
   g_s \to \epsilon^{3/4}\\
   \theta =\frac{1}{B}, \quad
   G =\frac{(2\pi\alpha' B)^2}{g}, \quad  
   \frac{1}{g_\mathrm{YM}^2}=\frac{g}{g_s2\pi\alpha^{'1/2}B} \quad :\ \mathrm{fixed} \label{eq:SW} 
\end{gather}
where $g$ is the closed string metric, $g_s$ is the string coupling and $B$ is strength of a background B-field on the D2-brane. This is the limit so that a noncommutivity is considerably larger than the string scale:
\begin{equation*}
   \Theta^2=\theta^{\mu\nu}\theta^{\alpha\beta}G_{\mu\alpha}G_{\nu\beta}=\frac{32\pi^4\alpha^{'4}B^2}{g^2} \gg \alpha^{'2}.
\end{equation*}
Therefore fields which have mass squared of order $1/\alpha'$ disappear, while those which have mass squared of order
\begin{equation*}
   \frac{1}{G\theta}=\frac{g}{(2\pi\alpha')^2B}
\end{equation*}
survive in the Seiberg-Witten limit. $-1/G\theta$ is noting but mass squared of the tachyon in our system. This is the reason why we can describe a dissolving D0-brane by using a noncommutative gauge theory in the Seiberg-Witten limit. 

\section{Rolling tachyon and dispersion of supergravity charge}\label{section:3}
In this section we study a process in which a D0-brane dissolves into a D2-brane by solving an equation of motion of the gauge field numerically, and evaluating supergravity charge distributions. At a glance one may think that time evolution of this system causes oscillation of gauge field modes around the potential bottom, rather than everlasting rolling toward it. Our results, however, indicate that the localized D0-brane disperses over all of the space on the D2-brane and the region in which the D0-brane disperses expands forever\footnote{Everlasting rolling toward a stable vacuum without oscillation around it can happen with infinitely many DOF which contribute to a tachyon condensation process. When a tachyon has a run-away potential, everlasting rolling can happen with a finite number of DOF, however there may exist singular behavior around the stable vacuum. }. This behavior seems natural from the viewpoint of dispersion of localized charge without any effect to confine it.

We numerically solve equations of motion with the Gauss low constraints of components of the operator $C$. Since it is difficult to deal with an infinite number of DOF we truncate the matrix to a finite size $(N+1) \times (N+1)$ for $N=100,1000,10000$. Our results show that gauge field modes roll down to the potential bottom, and then sudden disturbance happens at a certain time. This disturbance happens since we adopt matrix size truncation. Matrix size truncation has a geometrical interpretation that the space on the D2-brane is ``cut off''. Recall that a function on the noncommutative plane corresponding to an operator $|n\rangle\langle m|$ is given by $f_{nm}$ in \eqref{eq:Laguerre} which is a polynomial of $(n+m)$-th order of $\hat{r}$ with a Gaussian factor $e^{-\hat{r}^2}$. This means that higher number state modes correspond to more smeared configurations of the gauge field in the noncommutative space. In fact we can write a fluctuation of the gauge field  in terms of functions on the noncommutative space as
\footnote{Take care that $\hat{r}$ and $\hat{\varphi}$ are defined as operators which satisfy $[\hat{r}^2,\hat{\varphi}]=i$. Figure \ref{fig:laguerre1} should be considered just as an intuitive one because it is not justified to ignore the factor $e^{-\hat{\varphi}}$. In general it is difficult to draw a graph of functions on the noncommutative plane since they are represented in terms of operators.} (see Figure \ref{fig:laguerre1})
\begin{equation*}
   -i\theta \delta A_z = \sum_{n=1}^\infty d_{n-2}(-1)^n 2\sqrt{n}e^{-\hat{r}^2}(2\hat{r}^2)^{-1/2}e^{-i\hat{\varphi}}L_n^{-1}(2\hat{r}^2).
\end{equation*}
Hence matrix size truncation can be interpreted as restriction of the space on the D2-brane to a finite size. Taking $N\to\infty$ limit the full plane is recovered (see Figure \ref{fig:cutoff}). Therefore when the dispersed configuration of the gauge field expand beyond the ``cut off'' region, disturbance or a phenomenon to say ``reflection'' happens. In the limit the matrix size goes to infinity, we expect no ``reflection'' happens, and gauge field configuration disperses forever in the noncommutative space.

\psfrag{f_n}{$(-1)^n 2\sqrt{n}e^{-r^2}(2r^2)^{-1/2}L_n^{-1}(2r^2)$}\psfrag{n=1}{$n=1,t_0$}\psfrag{n=2}{$n=2,d_0$}\psfrag{n=12}{$n=12,d_{10}$}\psfrag{r}{$r$}
\EPSFIGURE{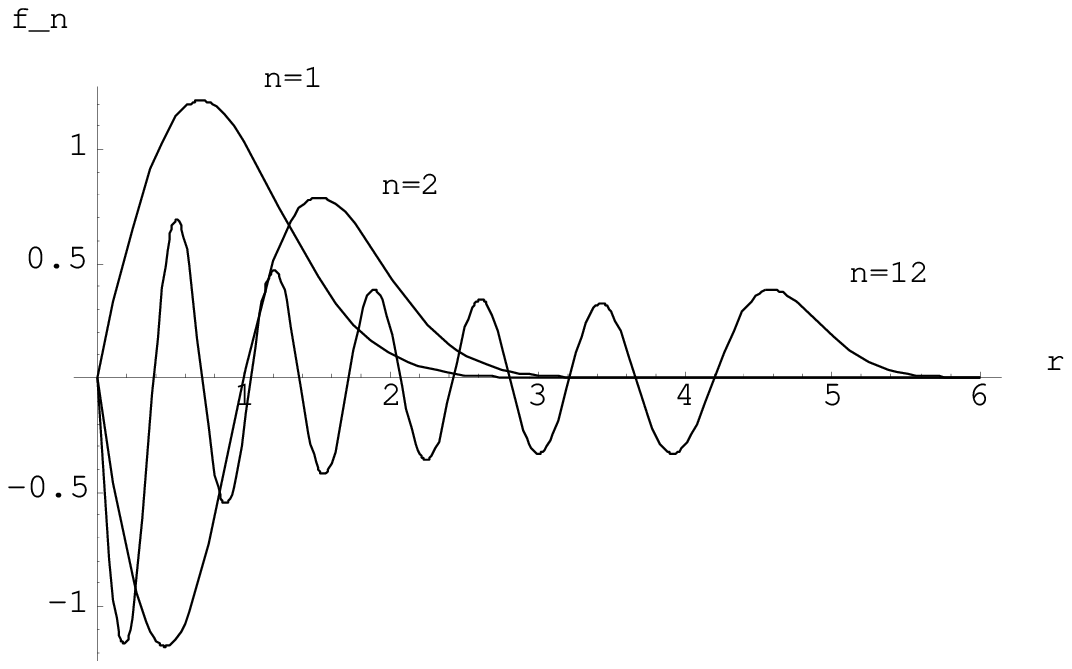,width=0.6\textwidth}{Mode expansion of noncommutative function\label{fig:laguerre1}}
\EPSFIGURE{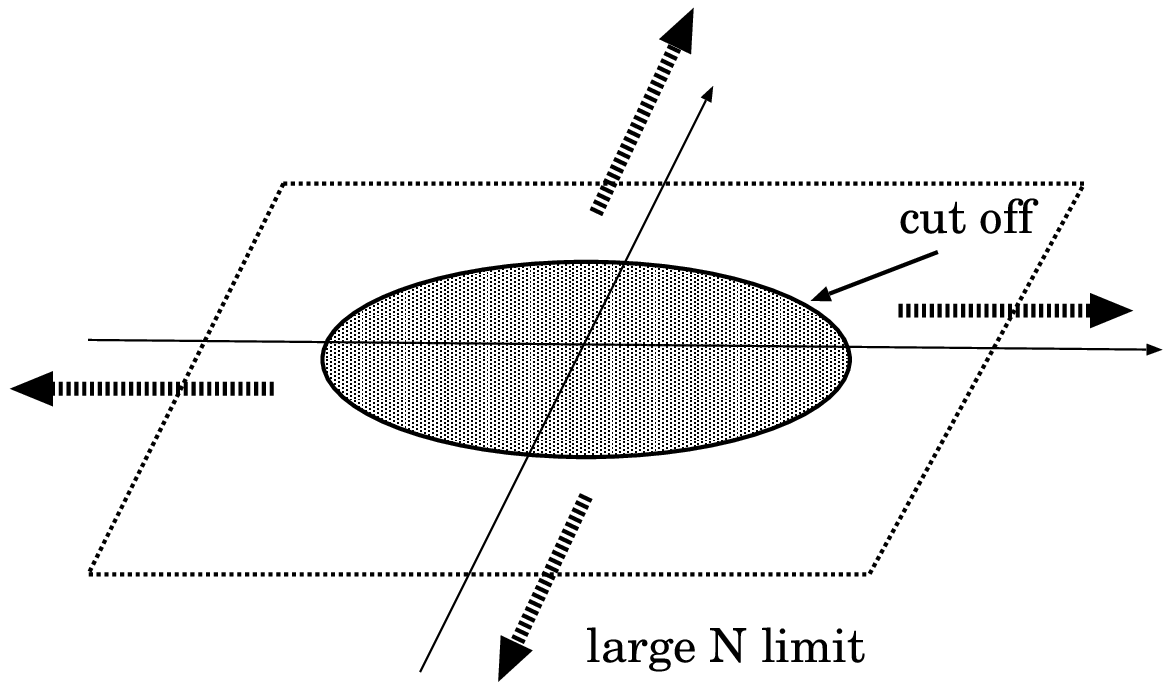,width=0.6\textwidth}{Geometrical interpretation of matrix truncation\label{fig:cutoff}}

Furthermore in this section we evaluate geometric distributions of D-brane charge, F-string charge, and energy density via formulas derived in Matrix theory. Gauge invariant observables can be extracted from a noncommutative gauge theory by using a matrix-valued configuration via these formulas. First we give prescriptions to calculate these formulas, and then we calculate charge distributions explicitly. Our results show that the initial localized D0-charge spreads out by turning on the tachyon mode, and disperses over the whole space on the D2-brane during the tachyon condensation process. The tachyon causes a negative infinity at the initial position of the D0-brane, and other mode gives a polynomial with a Gaussian factor of $r$ to D0-charge and energy density distributions. This negative divergence is expected to cancel out the initial singularity of D0-charge and energy density. D2-charge is not affected by any fluctuations, and a non-zero distribution of F-charge density without a singularity arises. We can verify that the charge conservation is satisfied kinematically \cite{0104036}. 

\subsection{Numerical calculation of time evolution}
In this subsection we numerically solve equations of motion with Gauss law constraints of components of $C$. Since it is difficult to deal with an infinite number of DOF, we truncate the matrix to a finite size $(N+1) \times (N+1)$. Then we have $t_0, d_i (i=0,1,\cdots,N-2)$ as modes to be considered. Substituting \eqref{eq:fluctuation} to \eqref{eq:EOM} and \eqref{eq:Gauss}, equations to be solved can be written as
\begin{equation*}
\begin{split}
  \partial_t^2 t_0&= -2t_0|t_0|^2+t_0|1+d_0|^2 \\
  \partial_t^2 d_0&= -2(1+d_0)|1+d_0|^2+(1+d_0)(|t_0|^2+|\sqrt{2}+d_1|^2) \\
                  &\ \ \vdots \\
  \scriptstyle \partial_t^2 d_{N-2}& \scriptstyle = -2(\sqrt{N-1}+d_{N-2})|\sqrt{N-1}+d_{N-2}|^2+(\sqrt{N-1}+d_{N-2})(|\sqrt{N-2}+d_{N-3}|^2+|\sqrt{N}+\Delta_{N-1}|^2)
\end{split}
\end{equation*}
\begin{equation*}
\begin{split}
  -\dot{t}^*_0t_0+t_0^*\dot{t}&=0 \\
   \dot{t}^*_0t_0-t_0^*\dot{t}-\dot{d}^*_0(1+d_0)+(1+d^*_0)\dot{d}_0&=0 \\
   &\  \vdots \\
  \scriptstyle \dot{d}^*_{N-3}(\sqrt{N-2}+d_{N-3})-(\sqrt{N-2}+d^*_{N-3})\dot{d}_{N-3}-\dot{d}^*_{N-2}(\sqrt{N-1}+d_{N-2})+(\sqrt{N-1}+d^*_{N-2})\dot{d}_{N-2} \scriptstyle &=0  
\end{split}
\end{equation*}
where $t$ has been re-defined as $t/\sqrt{G\theta}$ to become dimensionless. $\Delta_{N-1}$ is not a dynamical variable but a constant parameter which specifies the way of truncation, or we consider the time evolution in the section of $d_{N-1}=\Delta_{N-1}$. It seems natural to chose $\Delta_{N-1}=0,\sqrt{N+1}-\sqrt{N}$ to make $d_{N-1}$ stay at the potential top/bottom. Note that we treat Gauss law constraints to full order of fluctuations $\delta C$ since we are interested in time evolution beyond the region in which $\delta C$ is infinitesimal. Gauss low constraints can be rewritten as 
\begin{align*}
   \mathrm{Im}t_0(t)&=\alpha \mathrm{Re}t_0(t) \\
   \mathrm{Im}d_i(t)&=\beta_i (\sqrt{i+1}+\mathrm{Re}d_i(t)) \quad i=0,1,\cdots,N-2
\end{align*}
where $\alpha, \beta_i$ are time independent constants which are determined uniquely by an initial condition. Thus we can concentrate on real parts of equations of motion. In what follows we consider the case of $\mathrm{Im}t_0=0, \mathrm{Im}d_0=0,\cdots$. In this case $t_0,d_i$ remains real. 

We set an initial condition as $t_0=1.0\times 10^{-6}, d_i=1.0\times 10^{-6}(i=0,\cdots,N-2)$ and solve equations of motion numerically by using the embedded Runge-Kutta Prince-Dormand (8,9) method keeping absolute error less than $1\times 10^{-6}$ or relative error less than $1\times 10^{-3}$ for N=100,1000,10000 and $\Delta_{N-1}=0,\sqrt{N+1}-\sqrt{N}$
\footnote{In the limit of infinite matrix size, it is not clear whether a given initial condition corresponds a configuration which belongs to the same topological sector as the unstable soliton. From this viewpoint it may seem good to turn on only the tachyon at $t=0$. We perform numerical calculations also for the initial condition $t_0=0.01, d_i=0$. Then qualitative behavior makes no difference between these two initial conditions.}.

Our results are plotted in Figure \ref{fig:100-top},\ref{fig:1000-top},\ref{fig:10000-top} for $\Delta_{N-1}=0$ and \ref{fig:100-bottom},\ref{fig:1000-bottom},\ref{fig:10000-bottom},\ref{fig:1000-rapid},\ref{fig:10000-rapid} for $\Delta_{N-1}=\sqrt{N+1}-\sqrt{N}$ respectively. These results figure out that at earlier time tachyon and lower number state modes\footnote{As is defined in section \ref{section:2}, we refer to $t_0$ in \ref{eq:fluctuation} as tachyon, $d_i$ of small $i$ as low number state modes, and $d_i$ of large $i$ as high number state modes. Note that $d_i$'s are not massive excited modes but massless ones of the gauge field.\label{footnote:1}} approach the potential bottom $t_0=1,d_i=\sqrt{i+2}-\sqrt{i+1}$, and as time goes higher number state modes sequentially roll down towards it. Considering that higher/lower number state modes correspond to more spread/localized configurations of the gauge field, we can interpret this as the process in which the gauge field configuration smears on the noncommutative plane. When the D0-brane disperses beyond the region on the D2-brane which is restricted to finite size because of matrix truncation, sudden disturbance or a phenomenon to say ``reflection'' happens. Since we have excluded distributions outside the domain except a Gaussian factor, propagation of the dissolving D0-brane over the ``boundary'' has no way other than ``reflection'' to inside. 

First we consider the case of $\Delta_{N-1}=0$. If we deal with $d_{N-1}$ as a dynamical variable, the system continues to roll down towards the stable vacuum, after it approaches a point $t_0=1,d_{i}=\sqrt{i+2}-\sqrt{i+1}\ (i \le N-2)$, since $V(d_{N-2}>0,d_{N-1}=0,d_{N}=0)$ has a negative derivative with respect to $d_{N-1}$. However, since we have fixed $d_{N-1}=\Delta_{N-1}=0$, there is no way other than ``reflection'' after the system approaches the point $t_0=,d_i=\sqrt{i+2}-\sqrt{i+1}\ (i \le N-2)$. 
Note that a potential minimum within the cross section of $d_{N-1}=0$ is a point $t_0=\sqrt{\frac{N^3}{N+1}-(N-1)N}$, $d_i=\sqrt{\frac{(N-i-1)N^2}{N+1}-(N-i-2)N}-\sqrt{N+1}\ (i \le N-2)$ of a potential height $V=\frac{1}{2G\theta(N+1)}$, but the point $t_0=1,d_{i}=\sqrt{i+2}-\sqrt{i+1}\ (i \le N-2)$ of a potential height $V=\frac{1}{2G\theta}$. 

On the other hand in the case of $\Delta_{N-1}=\sqrt{N+1}-\sqrt{N}$, $d_{N-2}$ feels a driving force and begins to roll down at $t=0$. Consequently $d_{N-2}, d_{N-3}, d_{N-4},\cdots$ roll down to the potential minimum in sequence and then rapidly oscillate around it (see Figure \ref{fig:1000-rapid},\ref{fig:10000-rapid}). This means that a wave of gauge field configuration propagating from the ``boundary'' to inside arises. When two waves, one from the original point to outside caused by the tachyon and one from the ``boundary'' to inside, collide each other, a sudden disturbance happens. In this case it may be more precise to call this phenomenon ``collision'' rather than ``reflection''. If $N$ is sufficiently large, the wave from the ``boundary'' becomes negligible and we do not need to distinguish ``reflection'' and ``collision'', or $\Delta_{N-1}=0$ and $\Delta_{N-1}=\sqrt{N+1}-\sqrt{N}$.

We can say that these results of numerical calculations are reliable until the disturbance happens. As $N$ becomes larger, size of the domain enlarges and thus occurrence of ``reflection'' delays. We expect that no ``reflection'' happens and the system causes everlasting rolling towards the stable vacuum in $N \to \infty$ limit. This process can be interpreted that the localized D0-brane disperses over the whole space on the D2-brane forever. 

\DOUBLEFIGURE[htb]{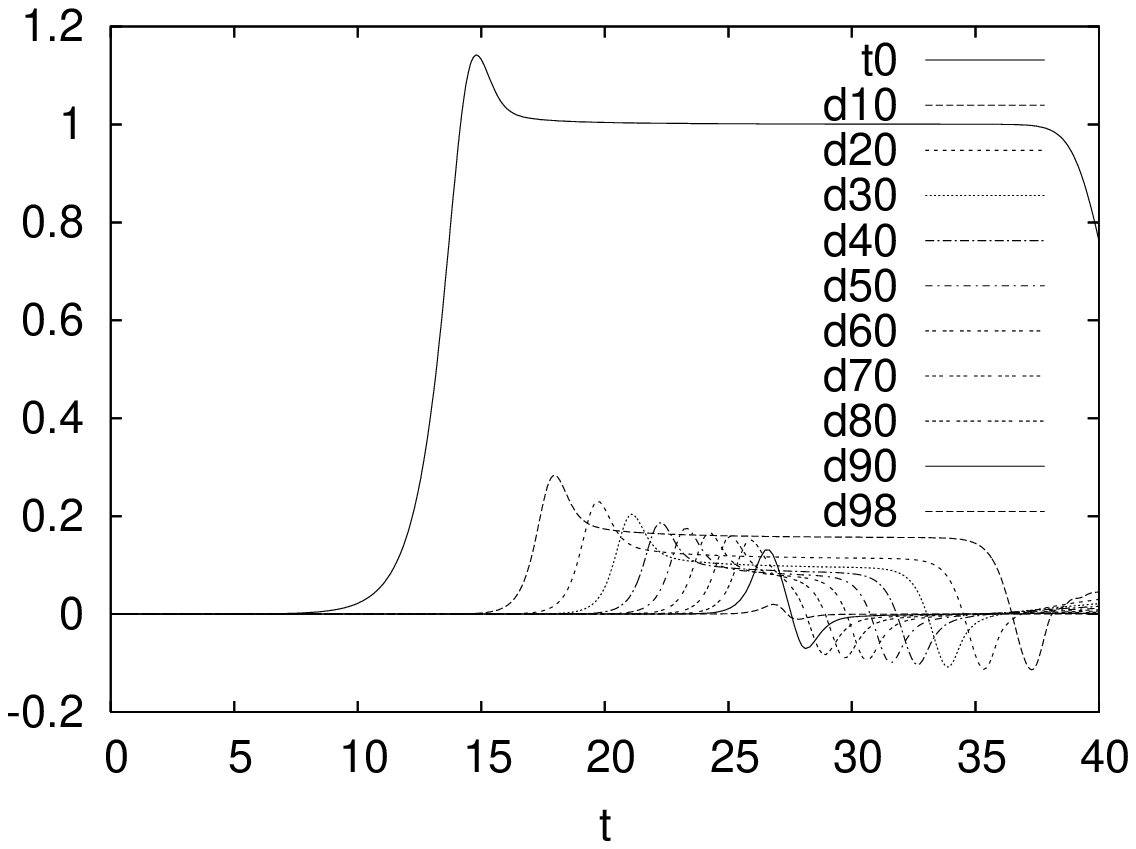,width=0.45\textwidth}{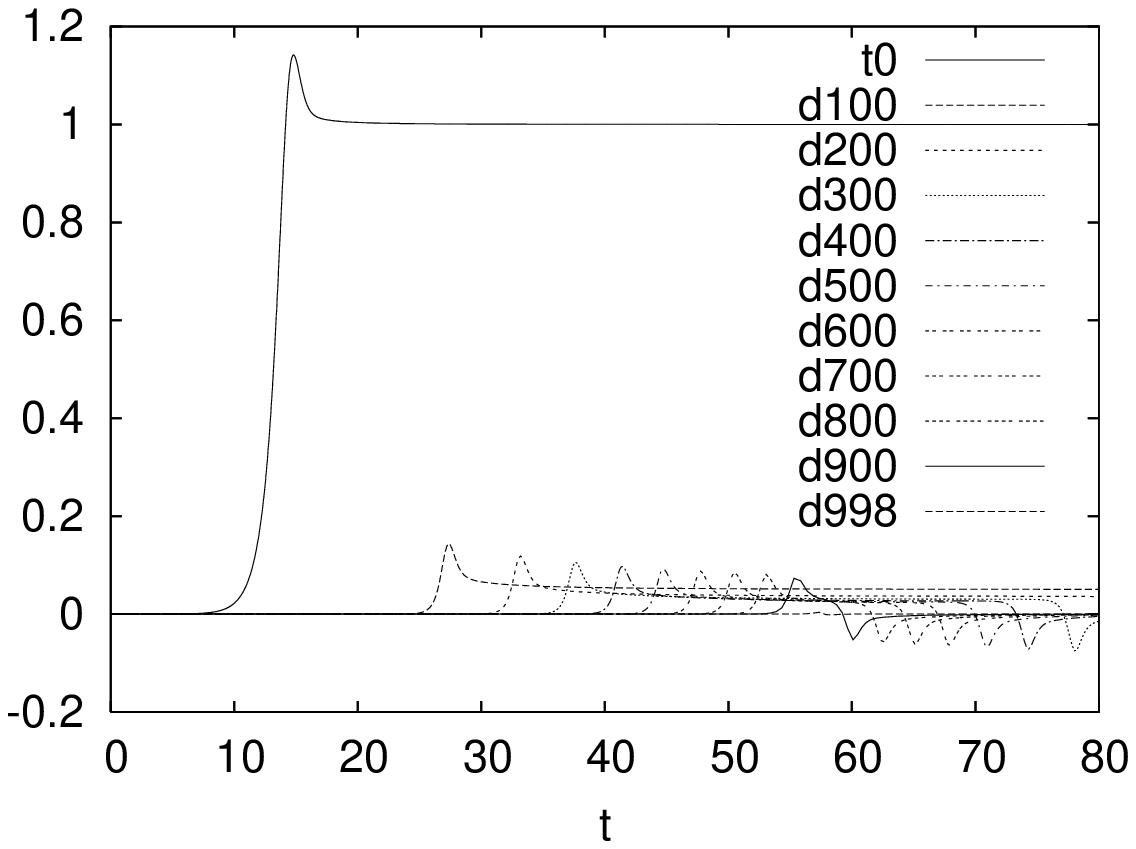,width=0.45\textwidth}{Results for $N=100,\Delta_{N-1}=0$\label{fig:100-top}}{Results for $N=1000,\Delta_{N-1}=0$\label{fig:1000-top}}

\EPSFIGURE[h]{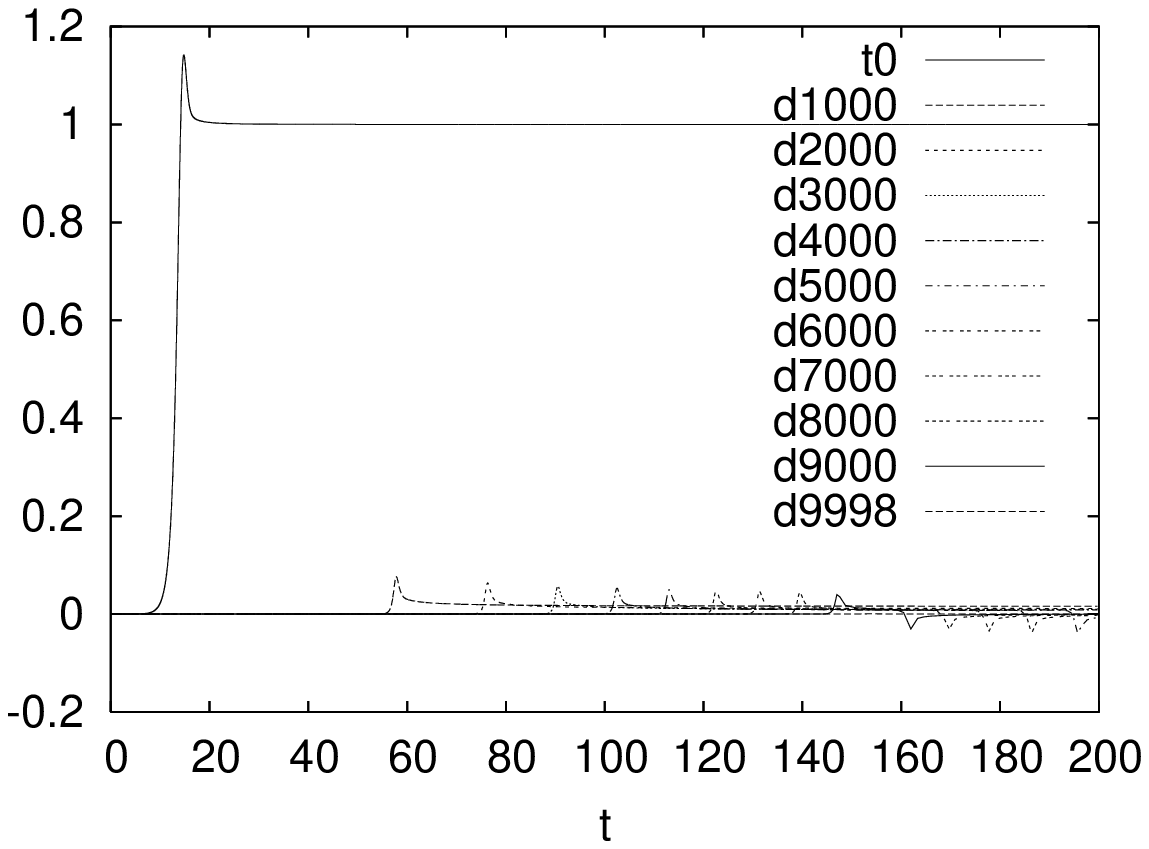,width=0.6\textwidth}{Results for $N=10000,\Delta_{N-1}=0$\label{fig:10000-top}}

\FIGURE[ht]{
\includegraphics[width=0.45\linewidth]{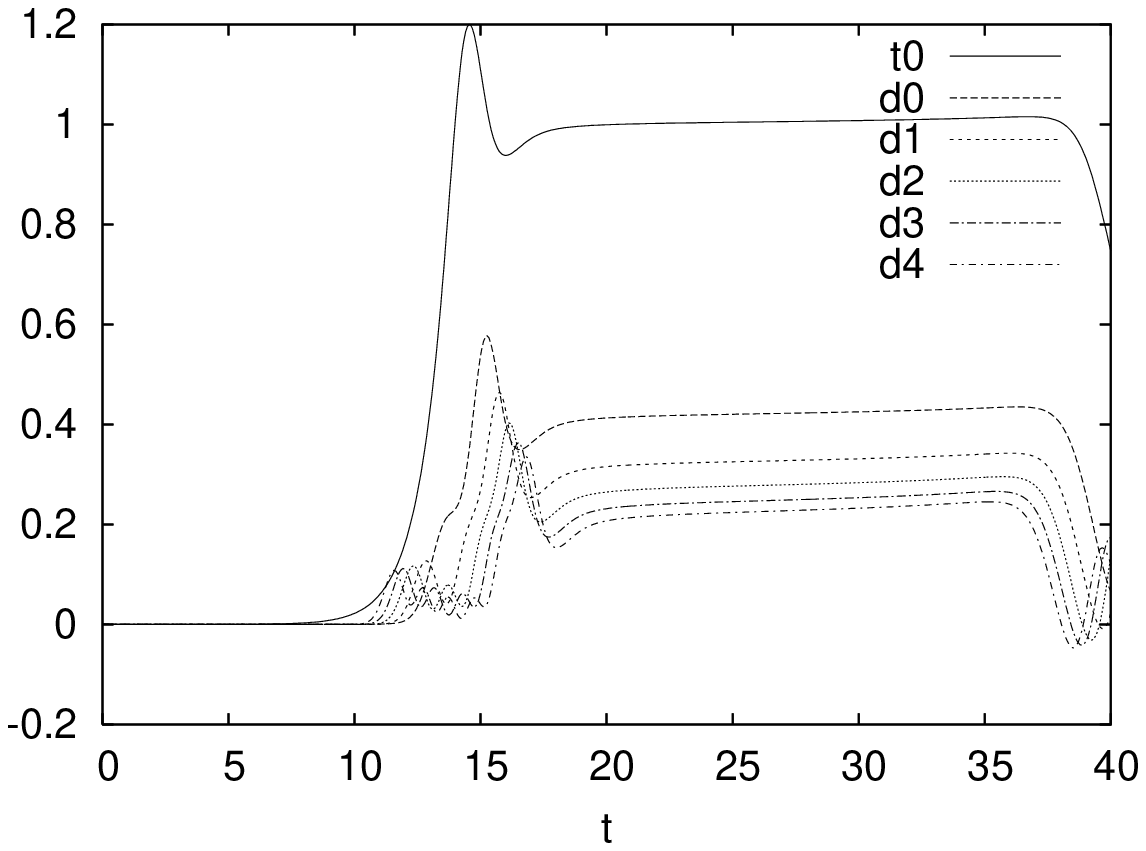}
\includegraphics[width=0.45\linewidth]{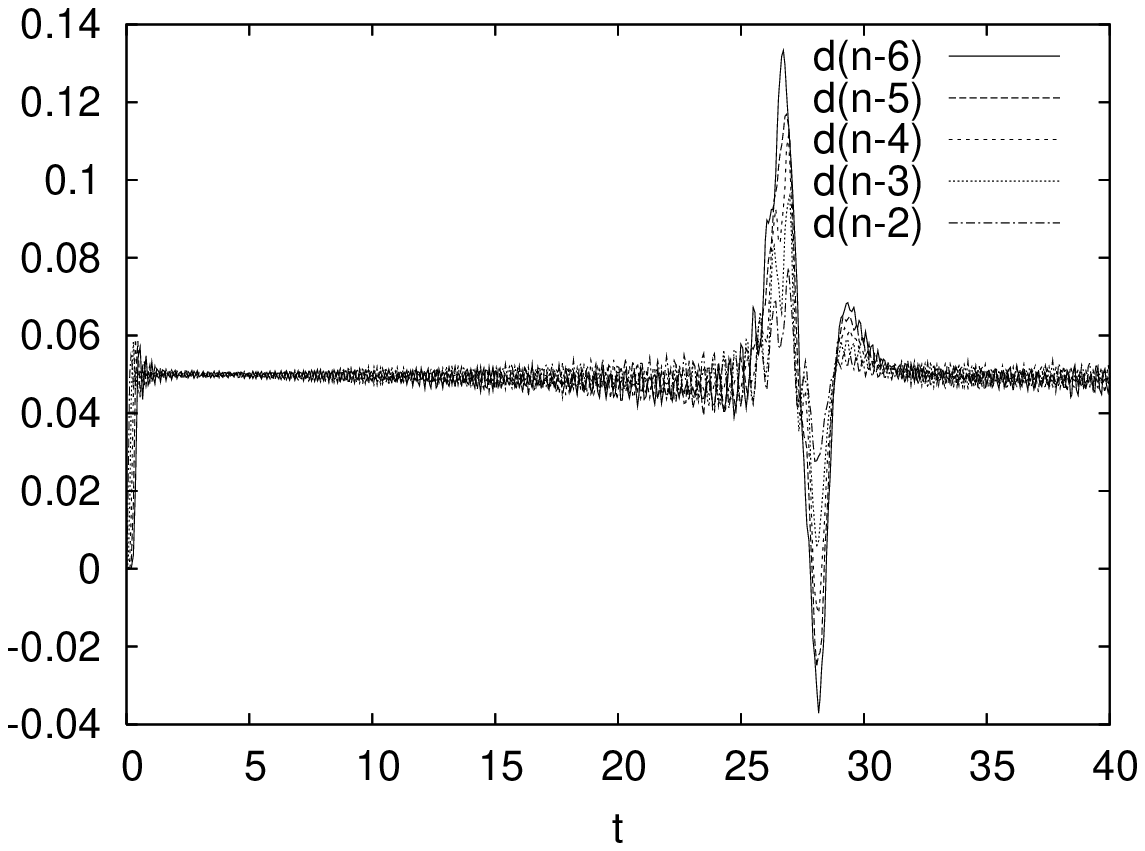}
\caption{Results for $N=100,\Delta_{N-1}=\sqrt{N+1}-\sqrt{N}$}\label{fig:100-bottom}
}

\FIGURE[ht]{
\includegraphics[width=0.45\linewidth]{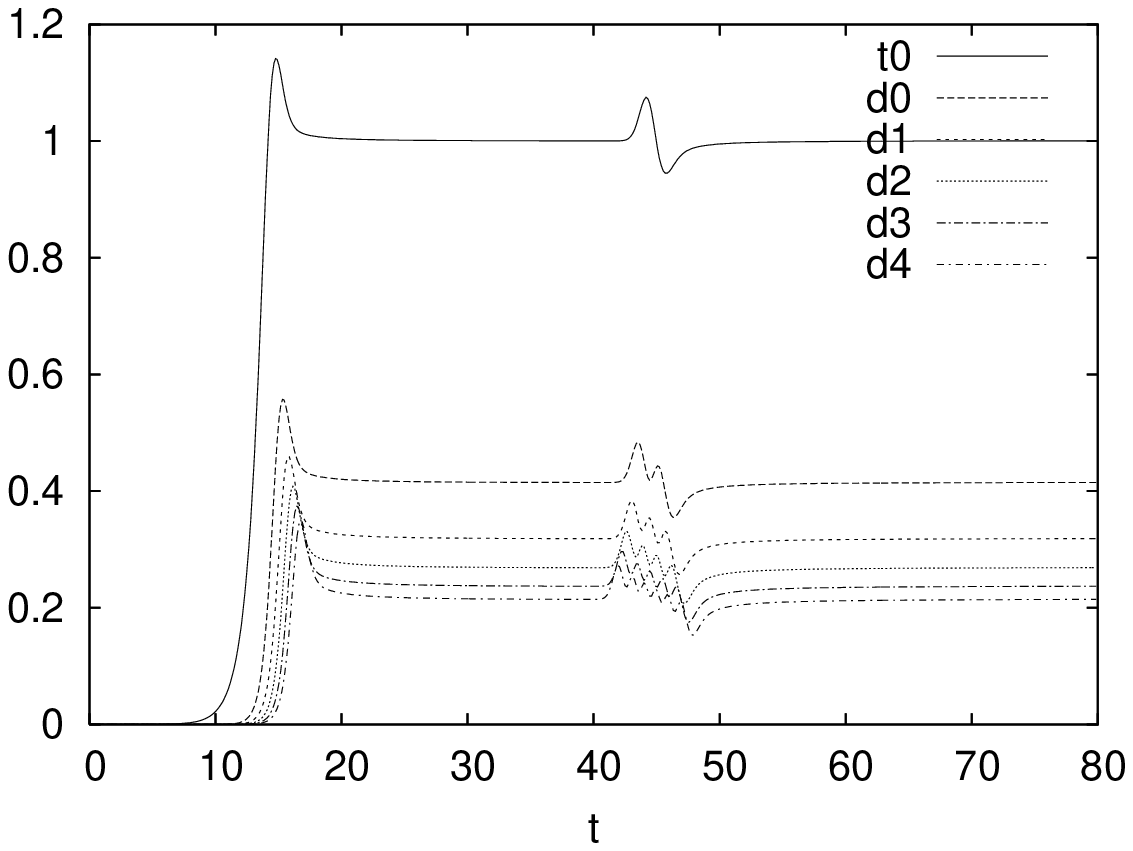}
\includegraphics[width=0.45\linewidth]{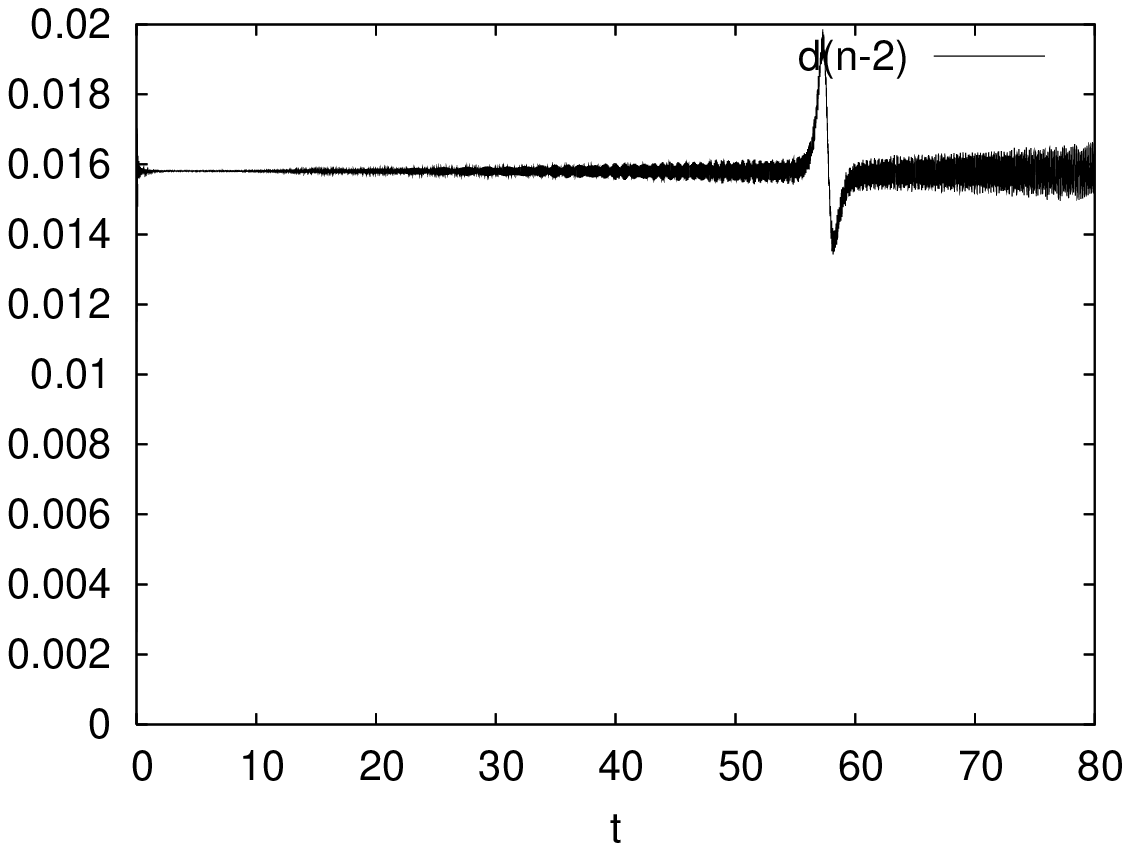}
\caption{Results for $N=1000,\Delta_{N-1}=\sqrt{N+1}-\sqrt{N}$}\label{fig:1000-bottom}
}

\FIGURE[ht]{
\includegraphics[width=0.45\linewidth]{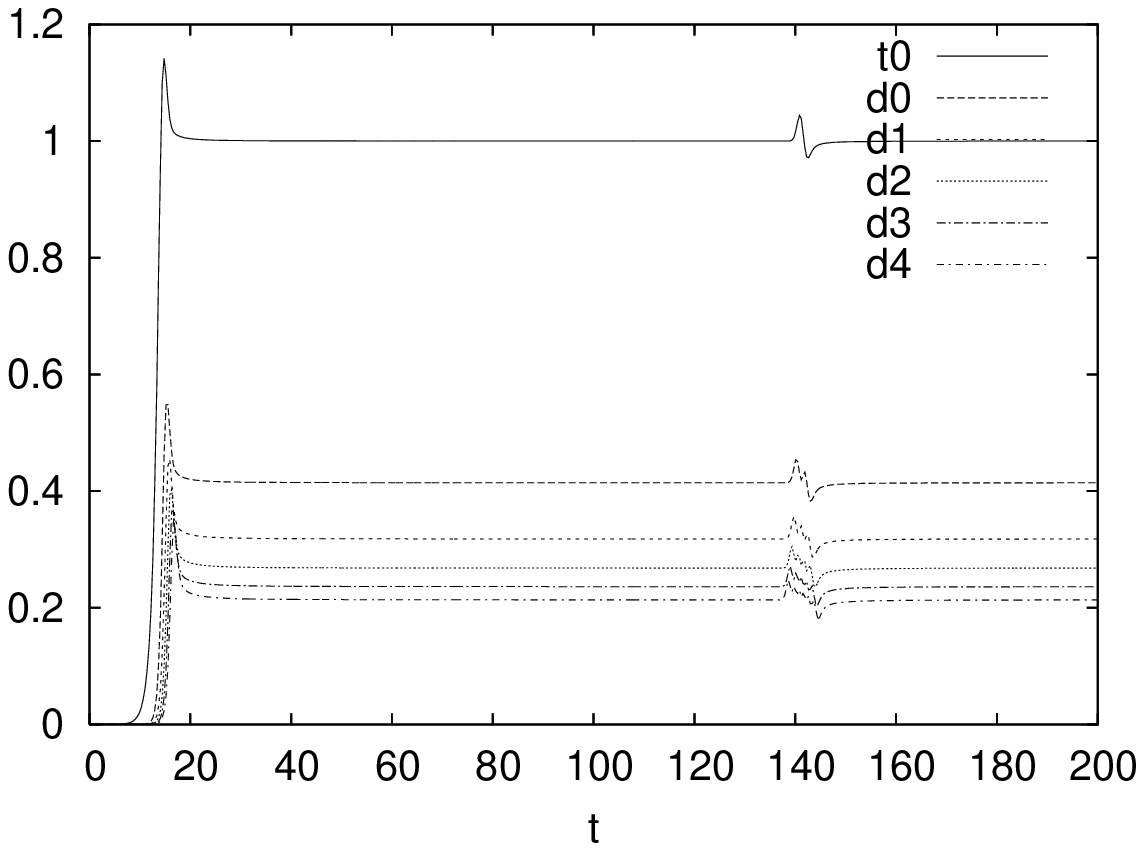}
\includegraphics[width=0.45\linewidth]{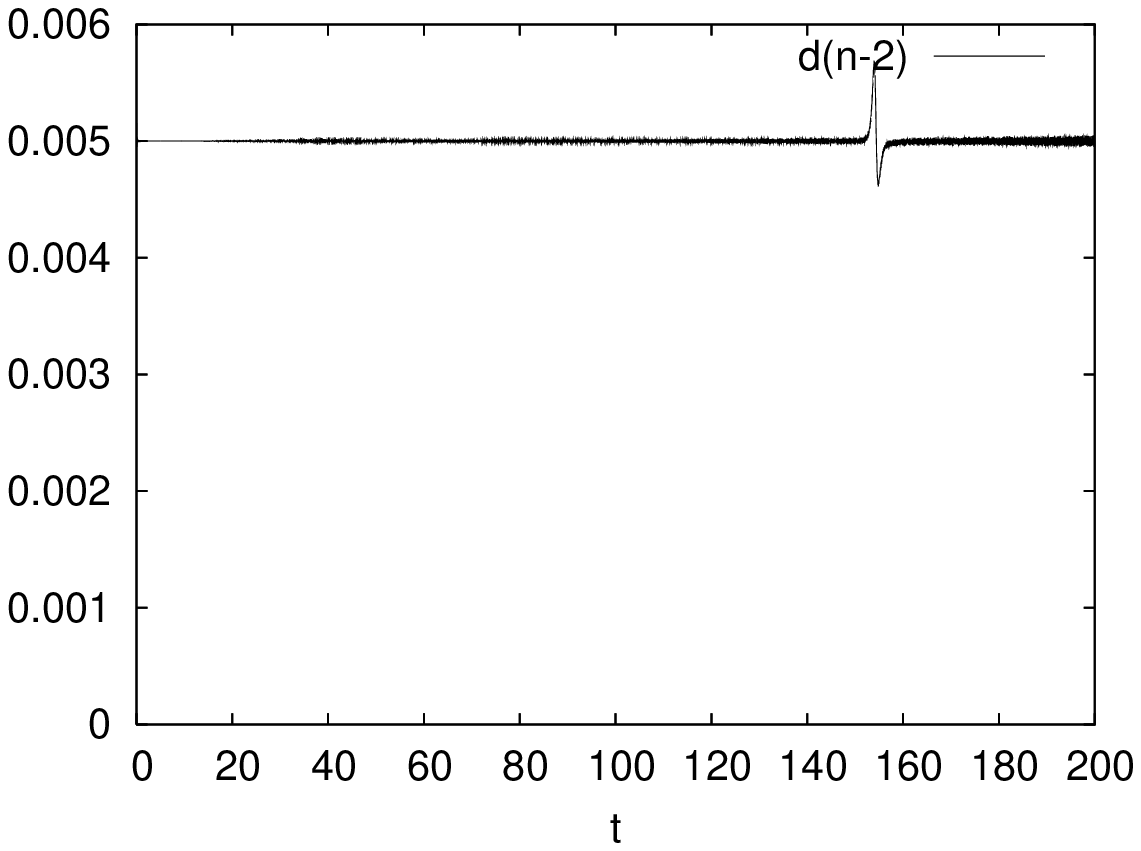}
\caption{Results for $N=10000,\Delta_{N-1}=\sqrt{N+1}-\sqrt{N}$}\label{fig:10000-bottom}
}


\DOUBLEFIGURE[htb]{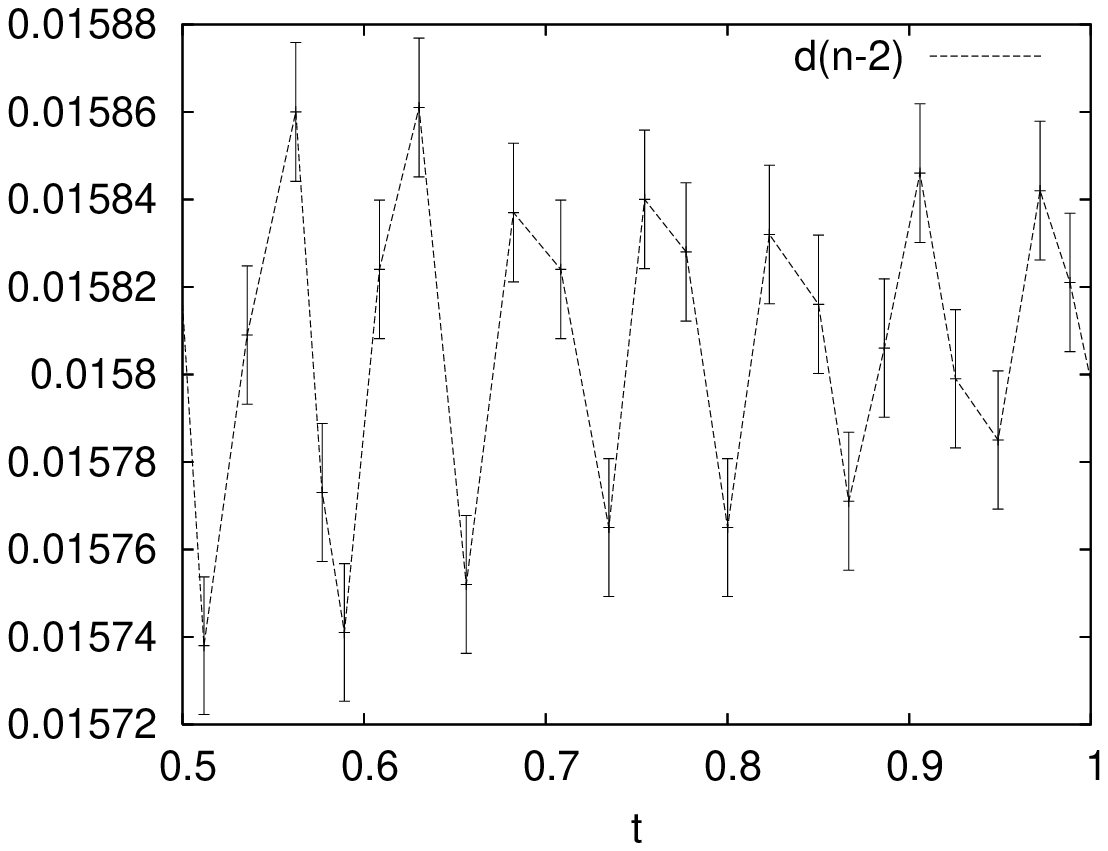,width=0.45\textwidth}{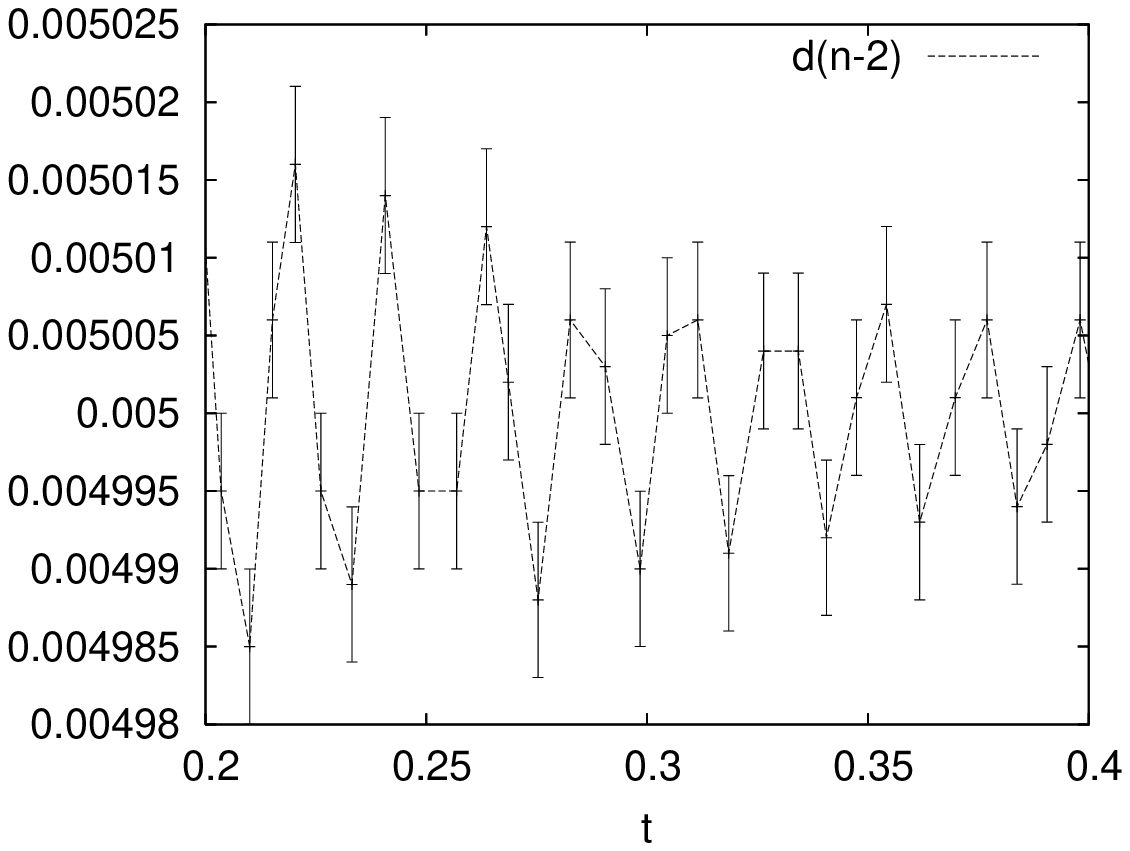,width=0.45\textwidth}{Rapid oscillation for $N=1000,\Delta_{N-1}=\sqrt{N+1}-\sqrt{N}$\label{fig:1000-rapid}}{Rapid oscillation for $N=10000,\Delta_{N-1}=\sqrt{N+1}-\sqrt{N}$\label{fig:10000-rapid}}

%
%
%
%
%
%

\newpage
\subsection{Supergravity charge distribution}
In oder to interpret the dissolving process geometrically, we calculate distributions of supergravity charge density. Formulas derived in Matrix theory \cite{9712185,9812239,0012218,0103124} are given in the momentum representation by 
\begin{align*} 
  \tilde{J}_0(k)&=T_{D0}\ \mathrm{Str}\left[e^{ikX}\right] \\
  \tilde{J}_2(k)&=\frac{T_{D0}}{2\pi\alpha'}\mathrm{Str}\left[-i\left[X^1,X^2\right]e^{ikX}\right] \\
  \tilde{I}^i(k)&=\frac{T_{D0}}{2\pi\alpha'}\mathrm{Str}\left[\left[X^i,X^j\right]\dot{X}_je^{ikX}\right] \\
  \tilde{T}(k)&=T_{D0}\ \mathrm{Str}\left[\left(\frac{1}{2}(\dot{X}^i)^2+\frac{1}{4(2\pi\alpha')^2}\left(i\left[X^i,X^j\right]+\theta^{ij}\right)^2+\left(1-\frac{\left(\theta^{ij}\right)^2}{4(2\pi\alpha')^2}\right)\right)e^{ikX}\right]
\end{align*}
where $\tilde{J}_0$ is D-charge, $\tilde{J}_2$ is D2-charge, $\tilde{I}^i$ is F-charge, $\tilde{T}$ is energy density, and $T_{D0}$ is tension of D0-brane. $\tilde{J}(k)=\frac{1}{(2\pi)^2}\int d^2x J(x) e^{ikx}$ means the Fourier transform of $J(x)$ \footnote{We denote noncommutative coordinates as $\hat{x}^i$ and commutative coordinates as $x^i$.}. $\mathrm{Str}$ is a symmetrized trace with the unit of symmetrization $X_i, \dot{X}_i,[X_i,X_j], k_iX^i$\footnote{Possibilities of commutator correction was also studies in the case of finite size matrix, or finite number D0-brane for various noncommutative geometry \cite{0401043}. In our case there is no such correction since size of representation matrices is infinite}.
Validity of the formula for energy-momentum tensor was checked from the viewpoint of string scattering amplitudes \cite{0012218,0104036}. A relation of Ramond-Ramond potentials to the Seiberg-Witten map was also investigated in \cite{0104036}. Our results show that initial singularities of localized D0-charge and energy density are resolved by turning on the tachyon. Contributions of all other modes give polynomials with Gaussian factor of the distance from the original D0-brane position. 

\EPSFIGURE[h]{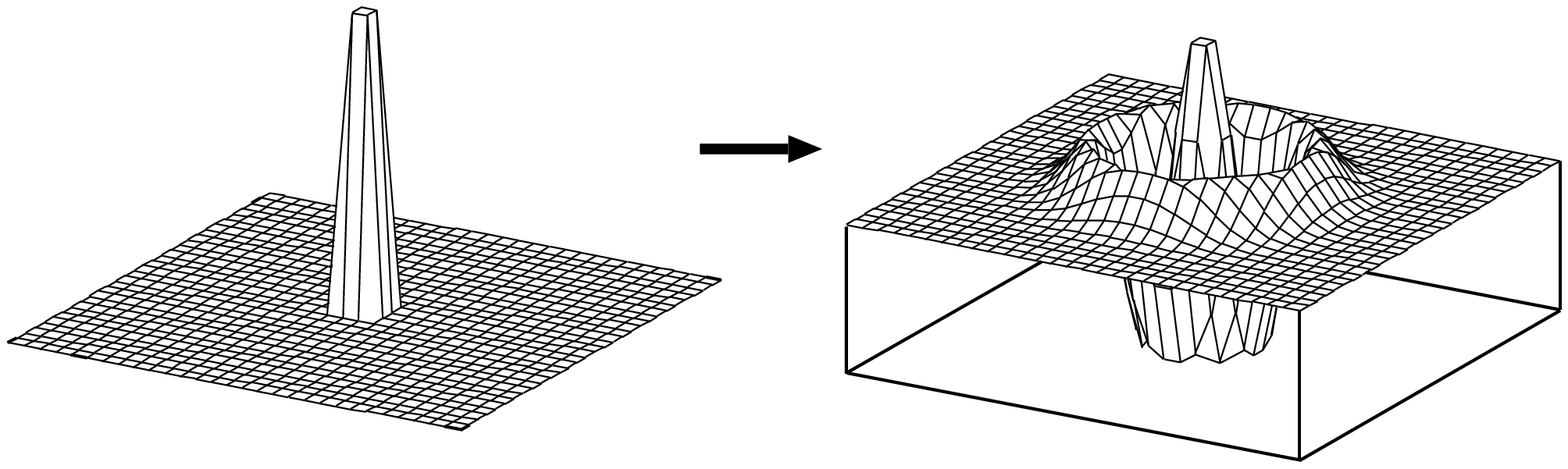,width=1.0\linewidth}{Resolution of charge density localization}\label{fig:fluctuation}

\subsubsection{Prescription to calculate charge distribution formulas}
Before explicit computations we give prescriptions to calculate charge density formulas. First we need to know how to take a symmetrized trace. Denoting operators which belong to the unit of symmetrization as $A_1,A_2,\cdots$, a symmetrized trace of the form $\mathrm{Str}A_1\cdots A_ne^{ikX}$ can be calculated \cite{0103124,0104036} by using equations
\begin{align*}
   \mathrm{Str}A_1e^{ikX}&=\mathrm{tr}A_1e^{ikx} \\
   \mathrm{Str}A_1A_2e^{ikX}&=\mathrm{tr}\int_0^1\!\!ds\ A_1e^{iskX}A_2e^{i(1-s)kX} \\
   \mathrm{Str}A_1A_2A_3e^{ikX}&=\mathrm{tr}\int_0^1\!\!ds_1 \int_0^{1-s_1}\!\!ds_2\ A_1e^{is_1kX}A_2e^{is_2kX}A_3e^{i(1-s_1-s_2)kX} + (A_2 \leftrightarrow A_3).
\end{align*}
When an operator includes only one $A$, the symmetrized trace becomes the ordinary trace of it due to the cyclic property of a trace. Otherwise we can symmetrize operator ordering by using parameter integrals. Proofs of these equations are shown in \eqref{eq:b1},\eqref{eq:b2}.   

In order to perform calculations further, we want to expand the operator $e^{ikX}$ in small fluctuations $\delta X^i$. Take care that $[X_0^i,\delta X^j] \ne 0$ and $e^{ik(X_0+\delta X)}\ne e^{ikX_0}e^{ik\delta X}$ in general. We can expand $e^{ik(X_0+\delta X)}$ with respect to an infinitesimal deviation $\delta X^i$ as
\begin{equation*}
\begin{split}
  e^{ik(X_0+\delta X)}
=e^{ikX_0}&+\int_0^1\!\!d\tau\ e^{ik\tau X_0}ik\delta X e^{i(1-\tau)kX_0}\\
&+\int_0^1\!\!d\tau_1 \int_0^{1-\tau_1}\!\!d\tau_2 e^{ik\tau_1 kX_0}ik\delta X e^{i\tau_2 kX_0}ik\delta X e^{i(1-\tau_1-\tau_2)kX_0} +\cdots.
\end{split}
\end{equation*}
Taking the trace of this we get
\begin{equation*}
   \mathrm{tr}e^{ikX_0+ik\delta X}=\mathrm{tr}e^{ikX_0}+\mathrm{tr}e^{ikX_0} ik\delta X +\frac{1}{2}\mathrm{tr}\int_0^1 \!\! d\tau \ e^{i \tau kX_0  }ik\delta X e^{i(1-\tau)kX_0}ik\delta X + \cdots.
\end{equation*}
Proof of these relations is give in \eqref{eq:b3},\eqref{eq:b4}.  

In this paper we consider a small fluctuation around the unstable soliton $X^i_0=U\hat{x}^iU^\dagger$. Thus we should know matrix components of an operator $e^{Uik\hat{x}U^\dagger}$ in the number operator representation. Considering an equation
\begin{equation*}
   e^{Uik\hat{x}U^\dagger}=1+\sum_{n=0}^\infty\frac{1}{n!}U(ik\hat{x})^nU^\dagger=Ue^{ik\hat{x}}U^\dagger+|0\rangle\langle 0|,
\end{equation*}
it is sufficient to know matrix components of $e^{ik\hat{x}}$. These components can be computed by making use of a coherent state $|\lambda \rangle =e^{-\frac{|\lambda|^2}{2}}e^{\lambda a^\dagger}|0\rangle$ where $\lambda$ is a complex number. A coherent state has properties
\begin{equation*}
  a |\lambda \rangle = \lambda |\lambda \rangle 
\quad
  \langle m | \lambda \rangle = e^{-\frac{|\lambda|^2}{2}}\frac{\lambda^m}{\sqrt{m!}} 
\quad
  1  = \frac{1}{\pi}\int\!\! d^2\lambda \ |\lambda \rangle \langle \lambda| 
\end{equation*}
where $d^2\lambda$ is an integral over the real and imaginary parts of $\lambda$. Note that coherent states are not orthonormal, but satisfy
\begin{equation*}
  \langle \lambda' | \lambda \rangle = e^{-\frac{|\lambda'|^2}{2}-\frac{|\lambda|^2}{2}+\lambda^{'*}\lambda}.
\end{equation*}
Then $\langle m |e^{ik\hat{x}} | n \rangle$ is calculated as
\begin{equation}
\begin{split}
   \langle m | e^{ik\hat{x}} | n \rangle 
&= \frac{1}{\pi} \int \!\! d^2 \lambda\ e^{\frac{\theta}{4}k^2}\langle m | e^{i\sqrt{\frac{\theta}{2}}(k_1-ik_2) a}|\lambda \rangle \langle \lambda | e^{i\sqrt{\frac{\theta}{2}}(k_1+ik_2) a^\dagger} | n \rangle   \\
&=\frac{e^{\frac{\theta}{4}k^2}}{\pi\sqrt{m!n!}} \int \!\! d^2 \lambda\ \lambda^m {\lambda^*}^n e^{-|\lambda|^2+i\sqrt{\frac{\theta}{2}}(k_1-ik_2) \lambda + i \sqrt{\frac{\theta}{2}}(k_1+ik_2) \lambda^*} \\
&=\frac{e^{ -\frac{\theta}{4}k^2 }}{\pi\sqrt{m!n!}}\int \!\! d\lambda_1 d\lambda_2 \  (\lambda_1+i\lambda_2)^m(\lambda_1-i\lambda_2)^n e^{-\left(\lambda_1-i\sqrt{\frac{\theta}{2}}k_1\right)^2-\left(\lambda_2-i\sqrt{\frac{\theta}{2}}k_2\right)^2},
\end{split} \label{eq:9}
\end{equation}
where we have used the Baker-Campbell-Hausdorff formula $e^{X+Y}=e^Xe^Ye^{-\frac{1}{2}[X,Y]}$ for $X,Y$ which are commutable with $[X,Y]$. We can see that the trace of $e^{ik\hat{x}}$ becomes a delta function as is expected:
\begin{equation*}
  \mathrm{tr}e^{ik\hat{x}}
=\sum_{n=0}^\infty \frac{1}{n!} \int \!\! d\lambda_1 d\lambda_2 \ |\lambda|^{2n} e^{-|\lambda|^2+i\sqrt{2\theta}(k_1\lambda_1+k_2\lambda_2)+\frac{\theta}{4}k^2} 
=\frac{2\pi}{\theta}\delta^2(k).
\end{equation*}
The numerical factor $2\pi/\theta$ is consistent with the identification of a trace of an operator $\mathrm{tr}\hat{O}$ with an integral over noncommutative coordinates $\int \frac{d^2 \hat{x}}{2\pi\theta} O(\hat{x}^1,\hat{x}^2)$. $\langle m | e^{ik\hat{x}} | n \rangle$ can be expressed also in the form of summation as 
\begin{equation*}
     \langle m | e^{ik\hat{x}} | n \rangle =\sum_{l=0}^{\mathrm{Min}(m,n)} \frac{\sqrt{m!n!}}{(n-l)!(m-l)!l!}\left(i\sqrt{\frac{\theta}{2}}\right)^{n+m-2l}(k_x+ik_y)^{n-l}(k_x-ik_y)^{m-l}e^{-\frac{\theta}{4}k^2}.
\end{equation*}
For given $m,n$ we can carry out an integral in \eqref{eq:9}. Results for several different values of $m,n$ are shown below.
\begin{equation*}
\begin{split}
   (e^{ik\hat{x}})_{00}&=e^{-\frac{\theta}{4}k^2} \\
   (e^{ik\hat{x}})_{10}&=e^{-\frac{\theta}{4}k^2}\textstyle i\sqrt{\frac{\theta}{2}}(k_1+ik_2) \\
   (e^{ik\hat{x}})_{20}&=e^{-\frac{\theta}{4}k^2}\textstyle \frac{-\frac{\theta}{2}(k_1+ik_2)^2}{\sqrt{2}} \\
   (e^{ik\hat{x}})_{30}&=e^{-\frac{\theta}{4}k^2}\textstyle \frac{-i\left(\frac{\theta}{2}\right)^{3/2}(k_1+ik_2)^3}{\sqrt{6}} \\
   (e^{ik\hat{x}})_{11}&=e^{-\frac{\theta}{4}k^2}\textstyle \left(1-\frac{\theta}{2}k^2\right)\\
   (e^{ik\hat{x}})_{21}&=e^{-\frac{\theta}{4}k^2}\textstyle \frac{-i\sqrt{\frac{\theta}{2}}(k_1+ik_2)\left(-2+\frac{\theta}{2}k^2\right)}{\sqrt{2}} \\
   (e^{ik\hat{x}})_{31}&=e^{-\frac{\theta}{4}k^2}\textstyle \frac{\frac{\theta}{2}(k_1+ik_2)^2\left(-3+\frac{\theta}{2}k^2\right)}{\sqrt{6}}\\
   (e^{ik\hat{x}})_{22}&=e^{-\frac{\theta}{4}k^2}\textstyle \frac{1}{2}\left(2-\frac{\theta}{2}4k^2+\frac{\theta^2}{4}k^4\right) \\
   (e^{ik\hat{x}})_{32}&=e^{-\frac{\theta}{4}k^2}\textstyle \frac{i\sqrt{\frac{\theta}{2}}(k_1+ik_2)\left(6-\frac{\theta}{2}6k^2+\frac{\theta^2}{4}k^4\right)}{2\sqrt{3}}\\
   (e^{ik\hat{x}})_{33}&=e^{-\frac{\theta}{4}k^2}\textstyle \frac{1}{6}\left(6-\frac{\theta}{2}18k^2+\frac{\theta^2}{4}9k^4-\frac{\theta^3}{8}k^6\right) \\
                 & \ \vdots .
\end{split}
\end{equation*}
Note also that a relation
\begin{equation*}
     (e^{ik\hat{x}})_{nm}={(e^{i(-k)\hat{x}})_{mn}}^*
\end{equation*}
is satisfied. 

In this paper we concentrate on the rotationally symmetric fluctuation \eqref{eq:fluctuation}. Furthermore we set $d_i=0$ for $i \ge N-1$ by hand to make $\delta C$ a trace class operator
\footnote{Note that $C=Ua^\dagger U^\dagger$ and $C=a^\dagger$ have different magnetic flux, or belong to different topological sectors. A soliton number is invariant under variations of a gauge field which are continuous and does not change asymptotic behavior at far distance. However the definition of ``continuous and falling off'' variation in terms of operators is not clear. It seems natural to impose bounded and trace class operators for such variations considering that $\mathrm{tr}[C,\bar{C}]$ should be treated as a topological quantity, and $\delta C_{nm}$ for large $n,m$ corresponds to a variation of the gauge field at far distance. Here we impose a variation matrices $\delta C$ to be effectively finite size, or $\delta C_{nm}=0$ for $n,m \ge N+1$. Otherwise it may change asymptotics at infinity of the gauge field and break topology of the soliton. Concerning topology in noncommutative field theories see also \cite{0106048,0010006,0102182,0105242}.}.
Take care that this is not matrix truncation we have used in numerical calculations, but we just ignore modes of high number state or equivalently configurations of the gauge field at far distance. Hence $\delta X^i, \delta \dot{X}^i, \delta [X^i,X^j], ik\delta X$ are effectively finite size, or have non-zero components only inside the submatrix of a finite size $(N+1) \times (N+1)$.
Now we are interested in a trace of the form $\mathrm{tr} \left[O'_1 A_1' \cdots O'_n A_n'\right]$, where 
 $A' \in \{ \dot{X}^i, ik\delta X, [X^1,X^2] \}$ and
 $O'  \in \{ e^{Ui k_a \hat{x}U^\dagger} \}$.
We have to pay attention to treatment of the operator $[X^1_0,X^2_0] = i\theta UU^\dagger$ since it has nonzero components outside the finite size submatrix of $(N+1) \times (N+1)$
in contrast to $A \in  \{ \delta X^i, \delta \dot{X}^i, \delta [X^i,X^j], ik\delta X \}$. To avoid difficulty of infinite size matrices, $O_1' [X^1_0,X^2_0] O_2'$ should be preliminary computed algebraically. By using properties of shift operators we have
\begin{equation*}
 e^{Uik_1\hat{x}U^\dagger} UU^\dagger e^{Uik_2\hat{x}U^\dagger}= Ue^{i(k_1+k_2)\hat{x}}U^\dagger.
\end{equation*}
$e^{iskX}UU^\dagger e^{i(1-s)kX}$ also can be simplified by using this relation as shown in \eqref{eq:b5}.
After all $\mathrm{tr} \left[ O'_1 A_1' \cdots O'_n A_n'\right]$ can be reduced to a trace of the form 
\begin{equation*}
   \mathrm{tr} \left[O_1 A_1 \cdots O_m A_m \right]
\end{equation*}
where
\begin{align*}
O & \in \{ e^{Uik x U^\dagger} , Ue^{ik' x}U^\dagger \} \\
A & \in \{ \dot{X}^i, \delta [X^i,X^j], ik\delta X \}.
\end{align*}
Since we can regard an operator of type $A$ as an effectively finite size matrix, it is sufficient to know components of $O$ inside the finite size submatrix in oder to compute $\mathrm{tr} \left[O_1 A_1 \cdots O_m A_m \right]$. Put it all together we are able to compute charge distributions with our prescriptions. Take care that $A_i$ should be a trace class operator.

Then we evaluate distribution of D0-charge, D2-charge, F-charge and energy density. In this section we concentrate on a contribution of the tachyon mode which includes divergence. Contributions of other modes give only polynomials with Gaussian factors of the distance from the original point as shown in \ref{append:c}. Detailed calculations are presented in appendix \ref{append:c}. 

\subsubsection{D0-charge}
D0-charge distribution is given by
\footnote{The Seiberg-Witten map in two noncommutative dimensions is given by
\begin{equation*}
   F_{12}(k)+\frac{1}{2\pi\theta}\delta^2(k)=\mathrm{tr}e^{ikX}
\end{equation*}
where $F_{12}$ is field strength in commutative variables. Thus we can read off field strength in the commutative representation from our results in this subsection.}
\begin{equation*}
   \tilde{J}_0(k)=T_{D0}\ \mathrm{tr}\left[e^{ikX}\right].
\end{equation*}
In what follows in this subsection we evaluate a distribution of D0-charge density for the unstable soliton $X^i=Ux^iU^\dagger$ with a fluctuation around it. At the potential top $X^i=Ux^iU^\dagger$ D0-charge is computed as 
\begin{equation*}
  \tilde{J}_0(k)= T_{D0}\ \mathrm{tr}\left[e^{Uik\hat{x}U^\dagger}\right]=T_{D0}\ \mathrm{tr}\left[ Ue^{ik\hat{x}}U^\dagger+|0\rangle\langle 0|\right]
        =T_{D0}\left(1+\frac{2\pi}{\theta}\delta^2(k)\right) .
\end{equation*}
The Fourier transform $J_0(x)=\int\frac{d^2k}{(2\pi)^2}\tilde{J}_0(k)e^{ikx}$ gives
\begin{equation*}
   J_0(x)= T_{D0}\delta^2(x)+\frac{T_{D0}}{2\pi\theta}, 
\end{equation*}
namely a localized D0-brane at $x^i=0$ and smeared D0-charge density $\frac{T_{D0}}{{2\pi\theta}}$ bounded on a D2-brane \cite{0303204}. Therefore we can interpreted the unstable soliton as a single D0-brane on a D2-brane in terms of charge density distribution. Strictly speaking what we have called a D2-brane so far is a bound state of smeared D0-branes of constant density and a D2-brane, however we abbreviate it by D2-brane. 

Turning on a fluctuation around the unstable soliton, we expect that the localized D0-brane spreads out and dissolves into the D2-brane to formulate constant D0-charge density. Expanding the D0-charge formula in a small fluctuation around $X^i=Ux^iU^\dagger$, we have  
\begin{multline*}
T_{D0}\ \mathrm{tr}\left[e^{Uik\hat{x}U^\dagger+ik\delta X}\right]=T_{D0}\left(1+\frac{2\pi}{\theta}\delta^2(k)\right)\\
+ T_{D0}\mathrm{tr}\left[ik\delta X e^{Uik\hat{x}U^\dagger} + \frac{1}{2} \int_0^1 \!\! d\tau \ ik\delta X  e^{Ui\tau k\hat{x}U^\dagger} ik\delta X  e^{Ui(1-\tau)k\hat{x}U^\dagger} +\cdots \right].
\end{multline*}
In this subsection we concentrate on a contribution of the tachyon to D0-charge distribution. All other modes give polynomials with Gaussian factors of the distance from the initial D0-brane position with integrals over parameters as shown in \eqref{eq:c1},\eqref{eq:c2}. After calculations the contribution of the tachyon $\tilde{J}_0(k)_{t_0}$ becomes
\begin{equation*}
 \tilde{J}_0(k)_{t_0} = 
\frac{T_{D0}}{2}\int_0^1 \!\! d\tau \ e^{-\frac{\theta}{4}k^2\tau^2}\left(-\theta k^2 \right)t_0^2
\end{equation*}
up to the second order of $t_0$. The Fourier transform of this gives
\begin{equation*}
\begin{split}
J_0(x)_{t_0}
&=T_{D0} \int_0^1\!\! d\tau \ \frac{4(x^2-\theta\tau^2)}{\pi\theta^2\tau^6} e^{-\frac{x^2}{\theta\tau^2}} t_0^2 \\
&=\begin{cases}
T_{D0}\frac{1}{\theta}\left\{\frac{1-\mathrm{Erf}(r)}{2\sqrt{\pi}r^3}+\frac{1+2r^2}{\pi r^2}e^{-r^2} \right\}t_0^2 \quad r\ne0 \\
-\infty\cdot t_0^2 \quad r=0
\end{cases} \quad r=\sqrt{\frac{(x^1)^2+(x^2)^2}{\theta}}
\end{split}
\end{equation*}
where Erf is the error function defined as $\mathrm{Erf}(x)=\frac{2}{\sqrt{\pi}}\int_0^x\!\! e^{-t^2}dt$, of which asymptotic behavior is given by
\begin{equation*}
 \mathrm{Erf}(r)=\begin{cases}
\frac{2}{\sqrt{\pi}}e^{-r^2}r\left(1+\frac{2r^2}{3}+\cdots\right) \qquad r \ll 1 \\
1-\frac{e^{-r^2}}{\sqrt{\pi}r}\left(1-\frac{1}{2r^2}+\cdots\right) \qquad r \gg 1.
\end{cases}
\end{equation*}
Therefore we can express D0-charge density as
\begin{equation*}
\begin{split}
   J(x)&=T_{D0}\delta^2(x)+J_(x)_{t_0}\\
       &=T_{D0}\frac{\partial^2}{\partial x^2}\left(\frac{1}{2\pi}\log(r)+\frac{1-\mathrm{Erf}(r)}{2\sqrt{\pi}r}t_0^2\right)
\end{split}
\end{equation*}
where we omit constant density $T_{D0}/2\pi\theta$. Around the original point and at far distance $J_0(x)_{t_0}$ acts like
\begin{equation*}
   J_0(x)_{t_0} \simeq \begin{cases} T_{D0} \frac{1}{2\theta\sqrt{\pi}r^3}t_0^2 \qquad r \ll 1 \\ T_{D0} \frac{2}{\theta\pi}e^{-r^2}t_0^2 \qquad r \gg 1 \end{cases}.
\end{equation*}
We can see that the D0-charge density distribution acts like the Gaussian distribution at far distance and diverges to positive infinity around the original point. The discontinuous behavior at $x=0$ can be also understood from the fact that an integral of $J_0(x)_{t_0}$ over $x$ vanishes:
\begin{equation*}
   \int_0^1\!\! d\tau \int\!\! d^2x \ \frac{2(x^2-\theta \tau^2)}{\pi\theta^2\tau^6}e^{-\frac{x^2}{\theta\tau^2}} = 0.  
\end{equation*}
This negative infinity at the original point is expected to cancel out the initial delta function singularity of D0-charge (see Figure \ref{fig:resolution}). Behavior of $J_0(x)_{t_0}/(T_{D0}t_0^2)$ around the original point is drawn in Figure \ref{fig:d0tachyon}. Take care that in this figure the divergence around the original point is cut off.
In order to see such resolution more precisely we should know contributions of the tachyon mode to all order. We can calculate such distributions up to a given order by following our prescriptions, however it is difficult to sum up all those terms to get a delta function which cancels out the initial delta function. We do not discuss this problem anymore in this paper. 

\psfrag{delta^2(x)}{$\delta^2(x)$}\psfrag{J_0(x)}{$J_0(x)$}\psfrag{-infinity}{$-\infty$}\psfrag{J_0(x)_{t0}}{$J_0(x)_{t_0}$}\psfrag{x1}{$x^1/\sqrt{\theta}$}\psfrag{x2}{$x^2/\sqrt{\theta}$}\psfrag{x}{$x$}
\EPSFIGURE[ht]{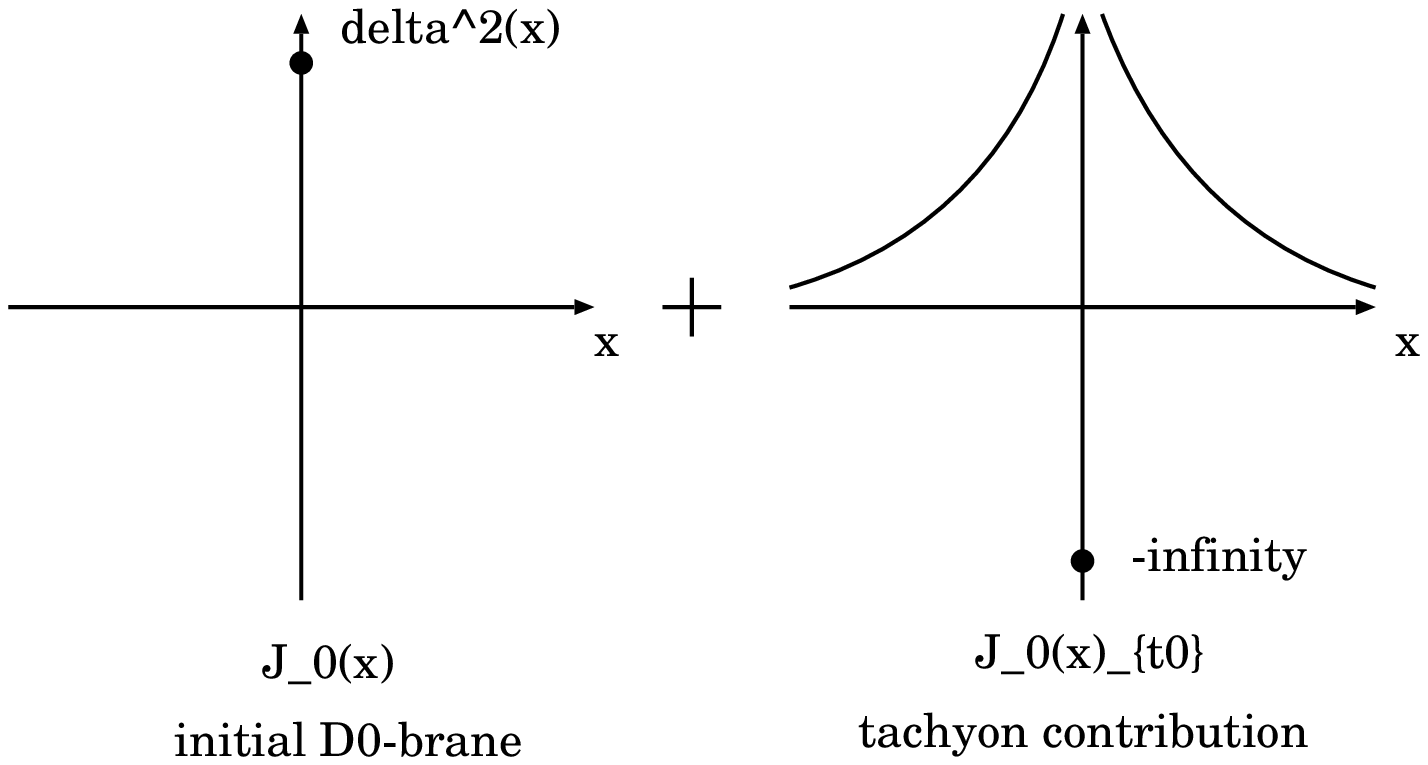,width=0.6\textwidth}{Resolution of D0-charge localization\label{fig:resolution}}
\EPSFIGURE[ht]{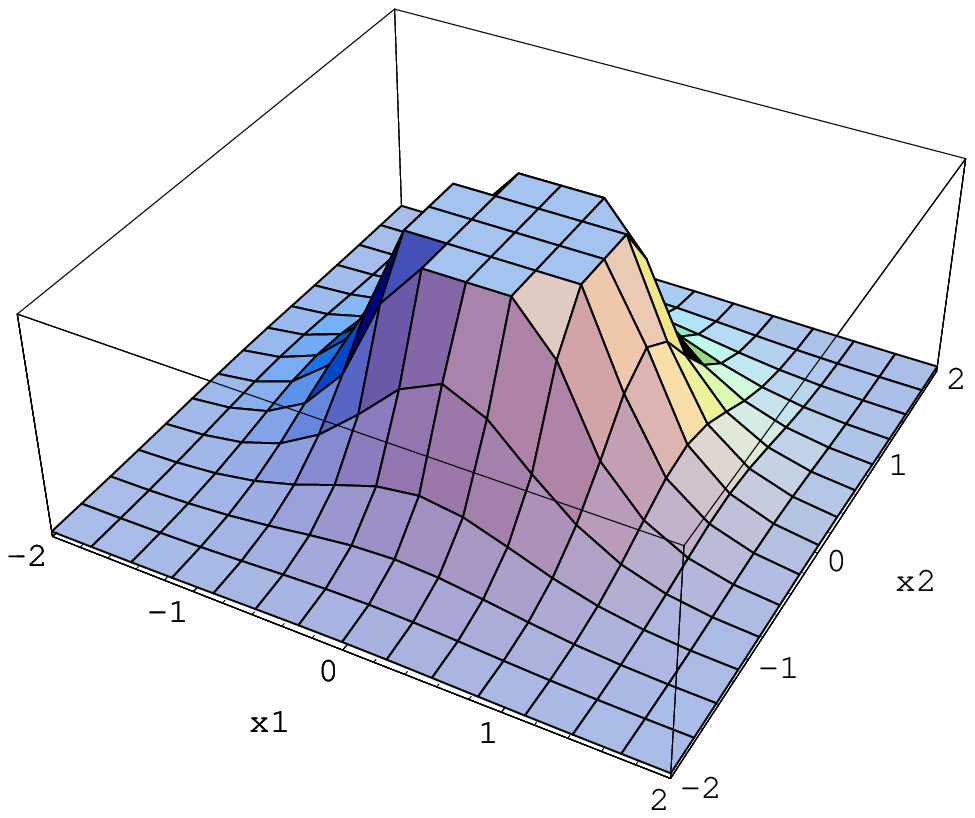,width=0.6\textwidth}{Smeared D0-charge around the original point\label{fig:d0tachyon}}

Next we verify the D0-charge conservation. Split the D0-charge density formula into an unstable soliton part and a fluctuation part as $\tilde{J}_0(k)=\tilde{J}_0^{(0)}(k) +\tilde{J}_0'(k)$, then the total charge conservation requires  
\begin{equation*}
   \int\!\! d^2x\ J'(x)=\int\!\!d^2x \int\!\!\frac{d^2k}{(2\pi)^2}\tilde{J}_0'e^{ikx}=\tilde{J}_0'(k=0)=0.
 \end{equation*}
This condition is satisfied from \eqref{eq:c1}. Thus total D0-charge is conserved during the dissolution process. Note that the conservation is satisfied kinematically \cite{0008075}.

\subsubsection{D2-charge}
D2-charge distribution can be written in the momentum representation as
\begin{equation*}
   \tilde{J}_2(k)=\frac{T_{D0}}{2\pi\alpha'}\mathrm{tr}\left[-i\left[X_1,X_2\right]e^{ikX}\right].
\end{equation*}
For the unstable soliton this formula gives
\begin{equation*}
   \tilde{J}_2(k)=\frac{T_{D0}\theta}{2\pi\alpha'}\mathrm{tr}\left[UU^\dagger \left(Ue^{ik\hat{x}}U^\dagger +|0\rangle \langle 0| \right)\right] = \frac{T_{D0}}{\alpha'}\delta^2(k)
\end{equation*}
and consequently D2-charge distribution is given by
\begin{equation*}
   J_2(x)=\int \!\!  \frac{d^2x}{(2\pi)^2}\ \frac{T_{D0}}{\alpha'}\delta^2(k) =T_{D2}.
\end{equation*}
Here $T_{D2}=T_{D0}/4\pi^2\alpha'$ is the D2-brane tension. This means the existence of a flat D2-brane in the 2+1 dimensional worldvolume as is expected. Next we evaluate a contribution of a fluctuation of the tachyon mode around the unstable soliton to the D2-charge distribution. After calculations presented in Appendix \ref{append:c} a contribution of the tachyon to D2-charge distribution in the momentum representation becomes 
\begin{equation*}
   \tilde{J}_2(k)_{t_0}=\frac{T_{D0}\theta}{2\pi\alpha'}\left\{(1-e^{-\frac{\theta}{4}k^2})-\int_0^1\!\!d\tau\ e^{-\frac{\theta}{4}(\tau^2+(1-\tau^2))k^2}\frac{\theta}{2}k^2 \left(1-\frac{k^2}{4}(1-2\tau)^2\right) \right\}t_0^2.
\end{equation*}
We can see that conservation of D2-charge $\tilde{J}'_2(k=0)=0$ is satisfied from \eqref{eq:c3}. The Fourier transform of this gives D2-charge distribution as
\begin{equation*}
\begin{split}
&   J_2(x)_{t_0}\\
&=\frac{T_{D0}\theta}{2\pi\alpha'} \left\{\delta^2(x)-\frac{e^{-\frac{x^2}{\theta}}}{\pi\theta}+\frac{e^{-\frac{x^2}{\theta}}}{\pi\theta}-\delta^2(x)\right\}t_0^2 =0.
\end{split}
\end{equation*}
Here we have used an relation
\begin{equation*}
    -\int_0^1 \!\! d\tau_1 \int_0^{1-\tau_1}\!\! d\tau_2 \ e^{-\frac{\theta}{4}(1-\tau_2)^2k^2}\frac{\theta}{2}k^2 e^{ikx}=\frac{e^{-\frac{x^2}{\theta}}}{\pi\theta}-\delta^2(k). 
\end{equation*}
We can say that there is a delta function in addition to the Gaussian distribution from the following equations:
\begin{align*}
   -& \int_0^1 \!\! d\tau_1 \int_0^{1-\tau_1}\!\! d\tau_2 \ e^{-\frac{\theta}{4}(1-\tau_2)^2k^2}\frac{\theta}{2}k^2 e^{ikx}=
\begin{cases}
\frac{e^{-\frac{x^2}{\theta}}}{\pi\theta} \qquad x \ne 0 \\
-\infty \qquad x=0
\end{cases}, \\
    & \int\!\! d^2x \int_0^1 \!\! d\tau_1 \int_0^{1-\tau_1}\!\! d\tau_2 \ e^{-\frac{\theta}{4}(1-\tau_2)^2k^2}\frac{\theta}{2}k^2 e^{ikx}=0.
\end{align*}
After all the tachyon mode has no effect on the D2-charge distribution. Contributions of other modes also vanish as shown in \eqref{eq:c4}. Therefore the D2-charge distribution kinematically remains 
\begin{equation*}
  J_2(x)=T_{D2}.
\end{equation*}

\subsubsection{F-charge}
F-charge distribution is given by
\begin{equation*}
\begin{split}
   \tilde{I}^1(k) & = \frac{T_{D0}g}{2\pi\alpha'}\mathrm{Str}\left[i[X^1,X^2]X^2\right]\dot{X}^2 \\
   \tilde{I}^2(k) & = \frac{T_{D0}g}{2\pi\alpha'}\mathrm{Str}\left[i[X^2,X^1]X^1\right]\dot{X}^1 \\
\end{split}
\end{equation*}
in the momentum representation. As is easily seen F-charge vanishes for the static solution. During the tachyon condensation process F-charge has a non-trivial distribution. A contribution of the tachyon to F-charge density becomes
\begin{align*}
\tilde{I}^1(k)_{t_0}&=
\frac{T_{D0}g}{2\pi\alpha'}\int_0^1 \!\! ds \int_0^1 \!\! d\tau\ e^{-\frac{\theta}{4}\theta^2(1-s\tau)^2k^2}(-ik_2)s t_0\dot{t}_0 \\
\tilde{I}^2(k)_{t_0}&=
\frac{T_{D0}g}{2\pi\alpha'}\int_0^1 \!\! ds \int_0^1 \!\! d\tau\ e^{-\frac{\theta}{4}\theta^2(1-s\tau)^2k^2}ik_1s t_0\dot{t}_0
\end{align*}
after calculations presented in appendix \ref{append:c}. We can see that total F-charge vanishes during the whole process since $\tilde{I}_i(k=0)=0$ is satisfied from \eqref{eq:c5},\eqref{eq:c6}. Take care that $\tilde{\vec{I}}$ takes the form of $\tilde{f}(k^2) \begin{pmatrix}-ik_2 \\ ik_1\end{pmatrix}$. The Fourier transform of $\tilde{I}$ gives F-charge distribution on the D2-brane as 
\begin{align*}
{I}^1(x)&=
\frac{T_{D0}g}{2\pi\alpha'}\int_0^1 \!\!ds \int_0^1 \!\! d\tau\ e^{-\frac{r^2}{(-1+s\tau)^2}}\frac{2}{\pi}\frac{sx_2}{(-1+s\tau)^4} t_0\dot{t}_0 \\
{I}^2(x)&=
\frac{T_{D0}g}{2\pi\alpha'}\int_0^1 \!\!ds \int_0^1 \!\! d\tau\ e^{-\frac{r^2}{(-1+s\tau)^2}}\frac{2}{\pi}\frac{s(-x_1)}{(-1+s\tau)^4} t_0\dot{t}_0.
\end{align*}
F-charge density $\vec{I}(x)$ takes the form of $f(r^2)\begin{pmatrix}x_2 \\ -x_1 \end{pmatrix}$, so it points the angular direction. We can see that $I^i(x=0)=0$ and there is no divergence at the original point. Behavior of $I^1(x)_{t_0}/(T_{D0}gt_0\dot{t_0})$ and $I^2(x)_{t_0}/(T_{D0}gt_0\dot{t_0})$ around the original point are plotted in Figure \ref{fig:f1tachyon} and \ref{fig:f2tachyon}. After all F-charge density formulates a non-trivial distribution without any divergence as shown in \eqref{eq:c7},\eqref{eq:c8}, and the total F-charge remains zero during the tachyon condensation. 

\DOUBLEFIGURE[htb]{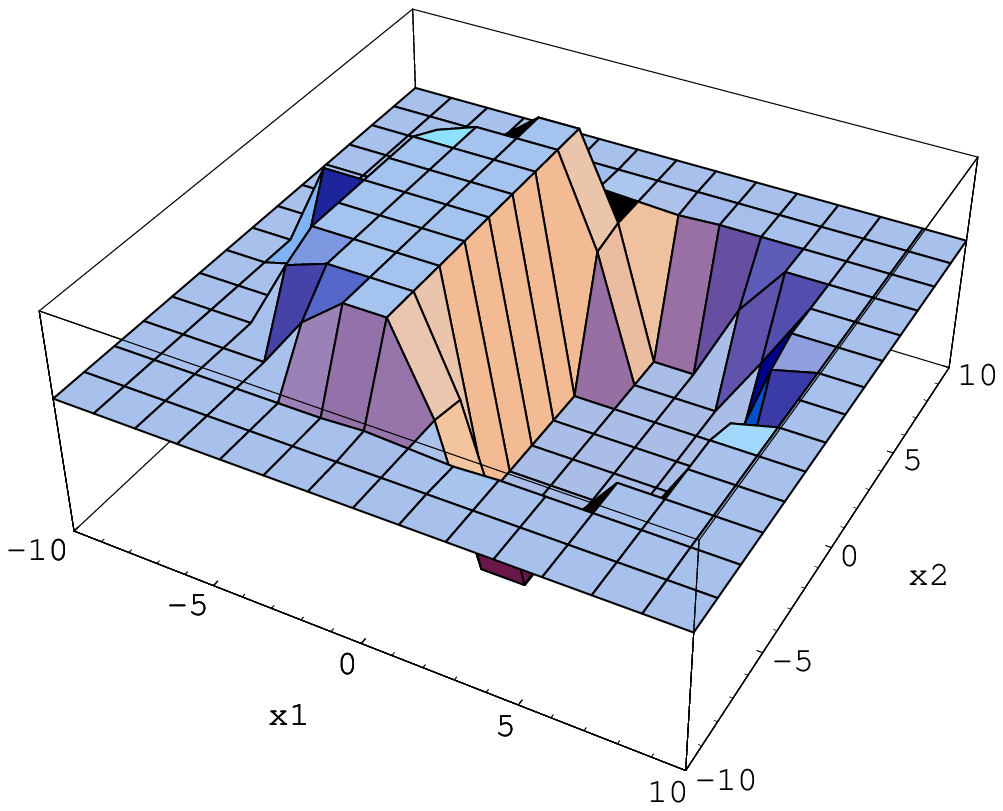,width=0.45\textwidth}{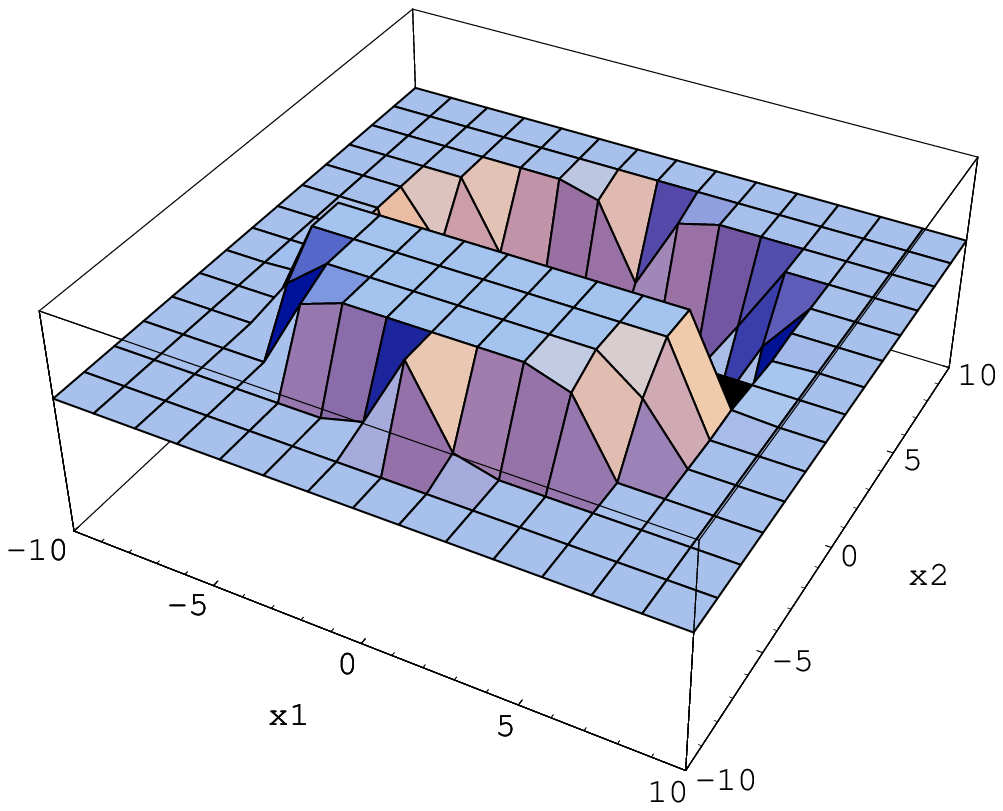,width=0.45\textwidth}{Behavior of $I^1(x)_{t_0}$\label{fig:f1tachyon}}{Behavior of $I^2(x)_{t_0}$\label{fig:f2tachyon}}

\subsubsection{Energy density}
Energy density distribution is given by
\begin{equation*}
\tilde{T}(k)=T_{D0}\ \mathrm{Str}\left[\left(\frac{g}{2}(\dot{X}^1)^2+\frac{g}{2}(\dot{X}^2)^2+1+\frac{ig\theta}{(2\pi\alpha')^2}-\frac{g^2}{2(2\pi\alpha')^2}\left[X^1,X^2\right]^2\right)e^{ikX}\right].     
\end{equation*}
This formula is different from the energy which is expected by the action \eqref{eq:action}
\begin{equation*}
E=\frac{2\pi}{\theta g_\mathrm{YM}^2} \mathrm{tr} \left[\frac{1}{2}(\dot{X}^1)^2+\frac{1}{2}(\dot{X}^2)^2 +\frac{1}{2G}+\frac{i}{G\theta}[X^1,X^2]-\frac{1}{2G\theta^2}[X^1,X^2]^2 \right]
\end{equation*}
by a constant shift $T_{D0}\left(1-\frac{g^2}{2(2\pi\alpha'B)^2}\right)$, which is identical to the energy of a D2-brane with constant magnetic field $B$ removing constant D0-charge density \cite{0009142}. Recall the parameter correspondence in the Seiberg-Witten limit \eqref{eq:SW}:
\begin{equation*}
   \frac{2\pi}{\theta g_\mathrm{YM}^2}=g T_{D0} \quad
   \frac{1}{2G\theta^2}=\frac{g}{2(2\pi\alpha')^2}.
\end{equation*}
The conservation of the total energy $\frac{d}{dt}\tilde{T}(k=0)=0$ is satisfied by using the equation motion. Take care that since we start from a point which deviates from the unstable soliton, $T'(k=0)$ is not a criterion of the energy conservation. In other words the total energy is conserved dynamically but not kinematically.
For the unstable soliton we have
\begin{equation*}
\begin{split}
&\tilde{T}(k)\\
=&T_{D0}\mathrm{tr}\left[e^{Uik\hat{x}U^\dagger}+\frac{i\theta g^2}{(2\pi\alpha)^2}UU^\dagger e^{Uik\hat{x}U^\dagger}+\frac{\theta^2 g^2}{2(2\pi\alpha')^2}\int_0^1\!\!ds\ UU^\dagger e^{Uisk\hat{x}U^\dagger} UU^\dagger e^{Ui(1-s)k\hat{x}U^\dagger}\right]\\
=&T_{D0}\left[1+\frac{2\pi}{\theta}\delta^2(k)\left(1-\frac{\theta^2 g^2}{2(2\pi\alpha')^2}\right)\right].
\end{split}
\end{equation*}
Then energy density is computed as
\begin{equation*}
   T(x)=T_{D0}\delta^2(x)+\frac{T_{D0}}{2\pi\theta}\left(1-\frac{\theta^2 g^2}{2(2\pi\alpha')^2}\right).
\end{equation*}
This is nothing but the energy of a localized D0-brane and D2-brane with constant D0-charge density. Next we turn on a fluctuation of the tachyon around the unstable soliton. After calculations presented in \ref{append:c} a contribution of the tachyon to the distribution of energy density becomes
\begin{equation*}
\begin{split}
&T(x)_{t_0}
=T_{D0}\left\{-\frac{2}{\pi\theta}\frac{g^2\theta^2}{2(2\pi\alpha')^2}e^{-r^2}+\int_0^1\!\!d\tau\ e^{-\frac{r^2}{(1-\tau)^2}}\frac{2(r^2-(1-\tau)^2)}{\pi\theta} \right.\\
&\phantom{T(x)_{t_0}=T_{D0} }\left.+\frac{g^2\theta^2}{2(2\pi\alpha')^2}\int_0^1\!\!d\tau_1 \int_0^{1-\tau_1}\!\!d\tau_2 \int_0^1\!\! ds\ e^{-\frac{r^2}{(1-s\tau_2)^2}}\frac{4s^2(r^2-(1-s\tau_2)^2)}{\pi\theta(1-s\tau_2)^6}\right\}t_0^2 \\
&\phantom{T(x)_{t_0}= }+T_{D0}\frac{g}{2}\int_0^1\!\!ds\ e^{-\frac{r^2}{(1-s)^2}}\frac{1}{\pi(1-s)^2} \dot{t}_0^2. +\cdots
\end{split}
\end{equation*}
as shown in \eqref{eq:c9},\eqref{eq:c10}. In this equation the first term is a Gaussian distribution and the second term is same as the one which appears in D0-charge density. We write down some equations in calculations of D0-charge density again:
\begin{equation*}
\begin{split}
&\int_0^1\!\!d\tau\ e^{-\frac{r^2}{\tau^2}}\frac{2(r^2-\tau^2)}{\pi\theta\tau^6} \\
&=\begin{cases}
-\infty \quad r=0 \\
\frac{1}{\theta}\left\{\frac{1}{4r^3}\sqrt{\pi}+\frac{1+2r^2}{2\pi r^2}e^{-r^2}-\frac{\mathrm{Erf}(r)}{4\sqrt{\pi}r^3} \right\}                
=\begin{cases}  \frac{1}{4\theta\sqrt{\pi}r} \qquad r \ll 1 \\  \frac{4}{\theta\pi}e^{-r^2} \qquad r \gg 1 \end{cases}.  
\end{cases}
\end{split}
\end{equation*}
The integrals in the third term is difficult to perform. The last term which includes $\dot{t}_0^2$ is computed as below.
\begin{equation*}
   \int_0^1\!\!ds\ e^{-\frac{r^2}{s^2}}\frac{1}{\pi s^2} = \frac{1-\mathrm{Erf}(r)}{2\sqrt{\pi}r}  
=\begin{cases} \frac{1}{2\sqrt{\pi}r} \quad r \ll 1 \\ \frac{e^{-r^2}}{2\pi r^2}  \quad r \gg 1  \end{cases}.
\end{equation*}
Behavior around the original point of the second term $\int_0^1\!\!d\tau\ e^{-\frac{r^2}{(1-\tau)^2}}\frac{2(r^2-(1-\tau)^2)}{\pi}$, the third term $\int_0^1\!\!d\tau_1 \int_0^{1-\tau_1}\!\!d\tau_2 \int_0^1\!\! ds\ e^{-\frac{r^2}{(1-s\tau_2)^2}}\frac{4s^2(r^2-(1-s\tau_2)^2)}{\pi\theta(1-s\tau_2)^6}$ and the last term $\int_0^1\!\!ds\ e^{-\frac{r^2}{(1-s)^2}}\frac{1}{\pi(1-s)^2}$ is plotted in Figure \ref{fig:energytachyon1}, \ref{fig:energytachyon2}, \ref{fig:energytachyondot}. Take care that in Figure \ref{fig:energytachyon1} the negative infinity at the original point is not plotted. Adding these terms all, the initial singularity of the energy density is expected to be resolved. After all our results show that localized energy density spread out by turning of the tachyon. As in the case of D0-charge, it is needed to calculate contributions of the tachyon to all order and sum up them all to study such resolution more precisely, however we do not discuss this problem anymore in this paper.

\DOUBLEFIGURE{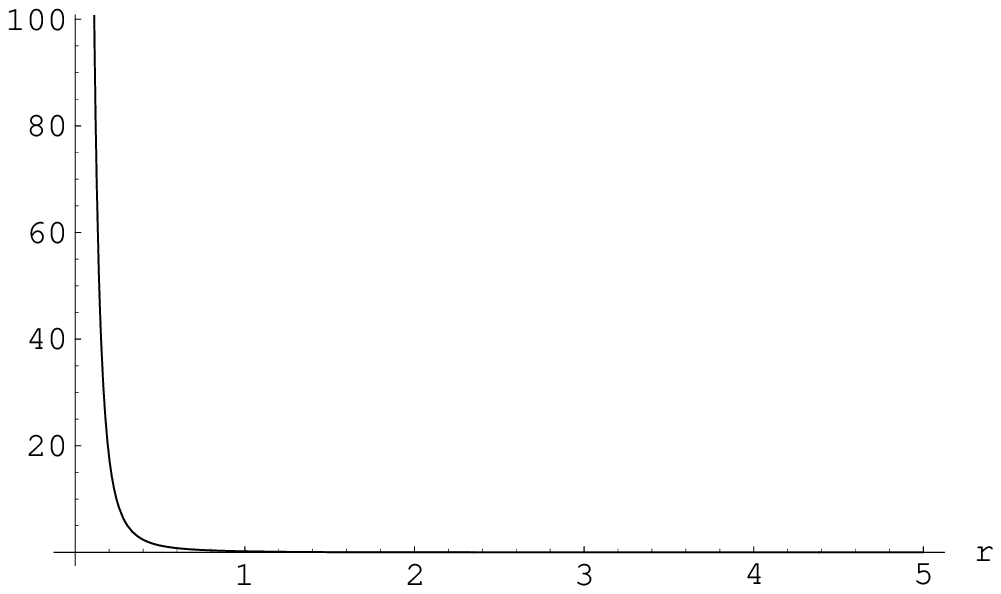,width=0.4\textwidth}{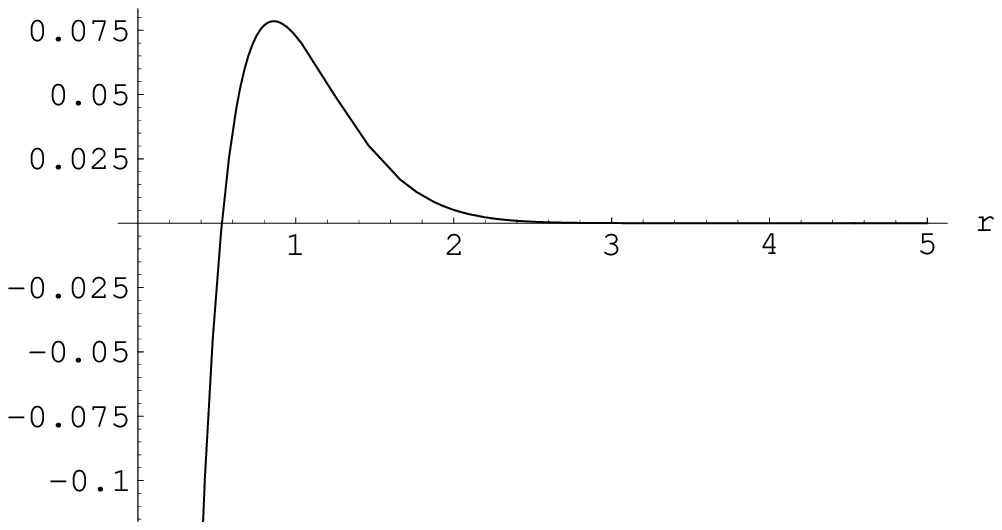,width=0.4\textwidth}{The second term\label{fig:energytachyon1}}{The third term\label{fig:energytachyon2}}
\EPSFIGURE{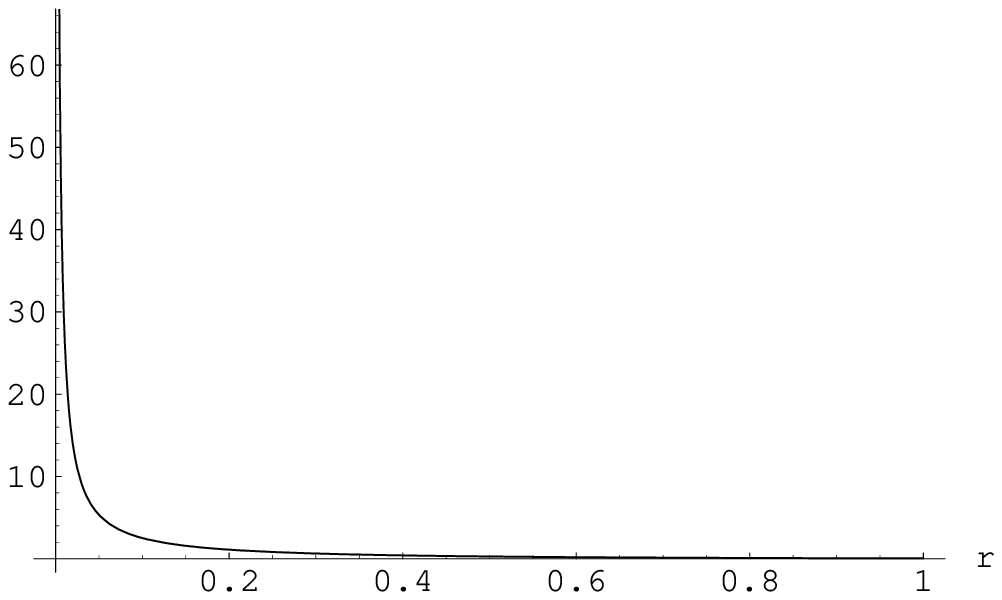,width=0.6\textwidth}{The last term\label{fig:energytachyondot}}

\subsection*{Charge distribution around the potential bottom}
In the remaining of this section we consider charge distributions around the potential bottom. 
When a fluctuation around the unstable soliton becomes large, it is not valid to evaluate charge distributions in expansion with respect to the fluctuation. Around the potential bottom we have $t_0=1, d_0=0.41\cdots, d_1=0.31\cdots, d_2=0.26\cdots,\cdots$. Thus we cannot ignore higher order terms of the tachyon in charge distributions. We can, however, evaluate charge distributions in expansion with respect to deviations from the potential bottom. Recall that $t_0,d_0,d_1,\cdots$ around the potential top metamorphose to fluctuations which correspond to $d_0,d_1,d_2,\cdots$ at the potential bottom. Hence our results of charge distributions around the potential top can be considered as those around the potential bottom except the tachyon mode and the localized soliton. This means that the tachyon condensation can be considered as the process in which D0-charge and energy density disperse on the D2-brane. 
 
\section{Conclusions and future directions}\label{section:4}
We have studied a dissolving D0-brane into a D2-brane with a constant B-field in the Seiberg-Witten limit. In this limit D0-branes on a D2-brane can be described as unstable solitons and its decay corresponds to a tachyon condensation towards a stable vacuum. In the rotationally symmetric case we have numerically solved the equation of motion with the Gauss law constraint of the gauge field. Our results indicate that the system rolls down asymptotically towards the stable vacuum. This can be interpreted as a process in which the localized soliton spreads out and disperses over all of noncommutative space on the D2-brane. Furthermore we have evaluated D0-charge, D2-charge, F-charge and energy density distribution on the D2-brane via formula derived in Matrix theory. We have provided prescriptions to calculate these formulas and evaluated them by following our prescriptions. Our results show that the initial singularities of D-charge and energy density are resolved by non-zero tachyon mode, and the region in which D0-brane disperses expands forever. 

Finally we discuss future directions of this work. Relaxing the rotational symmetry and turning on transverse scalars can be done straightforwardly. Taking account of fermionic fields is needed to study restoration of the supersymmetry. A quantum analysis of this system is also an important problem. 

One of the most interesting problem relating to this paper is about closed string emissions from a dissolving D0-brane. In the case of annihilation of a non-BPS D-brane, the initial energy is expected to be converted completely into a collection of closed strings \cite{0303139}. Then what is the remnant after D0-brane dissolution? It seems that we cannot deal with emissions of the energy to massive closed string modes in the zero slope limit. Even in the zero slope limit it is not easy to study the closed string emission from a time dependent D-brane and its backreaction. An answer to this problem will give an insight to the role of open and closed DOF in D-brane dynamics.  

Another interesting problem is to construct a dissolving D0-brane boundary state. It is shown that a higher dimensional D-brane can be constructed out of infinitely many D0-branes and anti-D0-branes in terms of boundary states. In particular a D2-brane is constructed out of infinitely many D0-branes \cite{9804163}. Recently explicit map from a D-brane with magnetic flux to a matrix configuration of D0-brane has been provided in a boundary state picture \cite{0501086}. We have studied in this paper a D0-brane which dissolves into a D2-brane from the viewpoint of a matrix-valued position of infinitely many D0-branes. Considering these all it seems possible to construct explicitly a boundary state which describes D0-brane dissolution. This idea suggests a method to study dynamics of some types of D-brane system. First interpret a configuration of higher dimensional D-branes as a collection of D0-branes. Then analyze dynamics of a matrix-valued configuration of such D0-branes. Finally re-interpret time evolution of the collection of D0-branes in terms of higher dimensional D-branes\footnote{This viewpoint is similar to the concept of K-matrix theory \cite{0101087,0108085,0212188}.}\footnote{Cocerning tachyon condensation in D0/D4-brane, see \cite{0502158}.}. A boundary state describing time evolution of this system can be constructed in terms of the collection of D0-branes, and a closed string emission can be evaluated by using the boundary state. Surely such an idea cannot be applicable to all D-brane dynamics, however, it will be a useful method to study some types of D-brane dynamics. 

\acknowledgments
I would like to thank to M. Kato and K. Hashimoto for useful discussions and comments.

\appendix
\newpage
\section{D-brane interpretation of noncommutative soliton}\label{append:a}
In this appendix we review spectrum matching between a noncommutative gauge theory in 2+1 dimensions and open strings in a D0/D2-brane system \cite{0009142,0011099}. To the first order of a fluctuations $\delta C$ the Gauss law constraint \eqref{eq:Gauss} gives
\begin{equation*}
  [\bar{C}_0,\delta C]+[C_0,\delta\bar{C}] =0,
\end{equation*}
or equivalently in terms of matrix components
\begin{equation*}
  -\sqrt{j-1}\delta C_{i,j-1}+\sqrt{i-1}\delta C^*_{j,i-1}+\sqrt{i}\delta C_{i+1,j}-\sqrt{j}\delta C^*_{j+1,i}=0. 
\end{equation*}
These constraints can be grouped with respect to $k=|i-j|$. A group of $k=0$ gives
\begin{subequations}
\begin{equation}
   \mathrm{Im} \delta C_{i+1,i}=0 \quad i=1,2,\cdots. \label{eq:gauss1}
\end{equation}
Each one equation from the other groups requires components of $W, T^\dagger$ to satisfy
\begin{equation}
   \sqrt{k-1}w_{k-2}+\sqrt{k}t_k=0 \quad k=1,2,\cdots. \label{eq:gauss2}
\end{equation}
\end{subequations}
\eqref{eq:gauss1} means that $\delta C_{i+1,i}=d_{i,i-1}$ should be real, but $\delta C_{10}=t_0$ remains complex. 
Quadratic terms in the potential of the action \eqref{eq:EOM} is
\begin{equation*}
\begin{split}
   V_2 & =\mbox{quadratic term in}\ \ \frac{1}{2G\theta}\mathrm{tr}\left([C,\bar{C}]+1\right)^2\\
     & = \frac{1}{G\theta}\mathrm{tr}\left\{\frac{1}{2}\left([a^\dagger,D^\dagger]+[a,D]\right)^2+(WW^\dagger-TT^\dagger)+(Wa-Ta^\dagger)(a^\dagger W^\dagger-aT)\right\}\\
     & =  \frac{1}{G\theta}\left\{\sum_{k=0}^\infty |\sqrt{k}w_{k-1}-\sqrt{k+1}t_{k+1}|^2-\sum_{k=0}^\infty (|w_k|^2-|t_k|^2)+\mathrm{tr}\left([a^\dagger,D^\dagger]+[a,D]\right)^2\right\}.
\end{split}
\end{equation*}
The complex number $A$ does not appear in the quadratic part of the potential and so we have two massless modes, which are identified to transverse scalars on a D0-brane. Components of the submatrix $D$ realize massless fluctuations of the gauge field on a D2-brane \cite{0007204,0502158}. Then we calculate a mass spectrum of $W,T^\dagger$. Relevant terms to matrix components $w_k,t_k$ can be regrouped as
\begin{gather*}
    \frac{1}{G\theta}\left\{-|t_0|^2+\sum_{k=1}^\infty |Y_k|^2 +\sum_{k=0}^\infty |Z_k|^2\right\}\\
   Y_k=\frac{1}{\sqrt{2k+1}}(\sqrt{k+1}w_{k-1}-\sqrt{k}t_{k+1}) \quad k=1,2,\cdots \\
   Z_k=\frac{1}{\sqrt{2k+1}}(\sqrt{k}w_{k-1}+\sqrt{k+1}t_{k+1}) \quad k=0,1,2,\cdots .
 \end{gather*}
Considering $Z_k$ is not dynamical due to Gauss law constraints \eqref{eq:gauss2}, $Y_k$ consists a mass spectrum
\begin{equation*} 
   m_k^2=\frac{2k+1}{G\theta} \quad k=1,2,\cdots
\end{equation*}
and $t_0$ turns out to be a complex tachyon with mass squared
\begin{equation*}
   m^2=-\frac{1}{G\theta}.
\end{equation*}
This mass spectrum is same as the one of open strings stretched between
a D0-brane and a D2-brane with a constant B-field in the Seiberg-Witten limit derived in terms of CFT \cite{9910263,0005283}
\begin{equation*}
   m_k^2=\frac{g(2k+1)}{(2\pi\alpha')^2B} \quad k=-1,1,2,\cdots
\end{equation*}
with the parameter correspondence in the Seiberg-Witten limit $\frac{1}{G\theta}=\frac{g}{(2\pi\alpha')^2B}$. After all we can identify
the unstable soliton in the noncommutative gauge field theory as a D0-branes on a
D2-brane with a constant background B-field from the viewpoint of spectrum matching. More rigorous derivation of the noncommutative gauge theory as an effective theory of this system based on CFT calculations was studied in \cite{0005283}.

\section{Proof of equations in prescription for charge distribution formulas}\label{append:b}
In this appendix we give proofs of some equations in our prescription to calculate charge density formulas. First we consider how to take a symmetrized trace \cite{0103124,0104036} of operators of the form $A_1\cdots A_n e^{ikX}$, denoting operators which belong to the unit of symmetrization as $A_i$. From the cyclic property of trace $\mathrm{Str}Ae^{ikX}$ becomes an ordinary trace $\mathrm{tr}Ae^{ikX}$, otherwise we need to symmetrize explicitly. We consider a symmetrized trace of the form $\mathrm{Str}ABe^{ikX}$, then we have
\begin{equation}
\begin{split}
  & \mathrm{Str}A_1A_2e^{ikX} \\
 &=\mathrm{tr}\sum_{n=0}^\infty\frac{1}{n!}\frac{n!}{(n+2)!}\sum_{\stackrel{0 \le i+j \le n}{i,j}} \!\!\left\{ (ikX)^i A_1 (ikX)^j A_2 (ikx)^{n-i-j} + (ikX)^i A_2 (ikX)^j A_1 (ikx)^{n-i-j}\right\}\\
 &=\mathrm{tr}\sum_{n=0}^\infty\frac{1}{(n+2)!}\sum_{j=0}^n \left\{(n-j+1)A_1 (ikx)^j A_2(ikX)^{n-j} +(n-j+1) A_1(ikx)^j A_2(ikX)^{n-j} \right\} \\
 &=\mathrm{tr}\sum_{p,q}\frac{p!q!}{(p+q+1)!}A_1 \frac{(ikX)^p}{p!} A_2 \frac{(ikX)^q}{q!}
  =\mathrm{tr} \int^1_0 \!\! ds \ A_1 e^{iskX} A_2 e^{i(1-s)kX} 
\end{split}\label{eq:b1}
\end{equation}
where we use an relation
\begin{equation*}
   \frac{p!q!}{(p+q+1)!}=\int_0^1 \!\! ds \ s^p(1-s)^q. 
\end{equation*}
In a similar way we can rewrite a symmetrized trace of the form $\mathrm{Str}A_1 A_2 A_3 e^ikX$ as
\begin{equation}
  \mathrm{Str}A_1A_2A_3e^{ikX}=\mathrm{tr} \int_0^1 \!\! ds_1 \int _0^{1-s_1} \!\! ds_2 \  A_1 e^{is_1kX} A_2 e^{is_2kX} A_3 e^{i(1-s_1-s_2)kX} +(A_2\leftrightarrow A_3) \label{eq:b2}
\end{equation}
where we use a relation
\begin{equation*}
   \frac{p!q!r!}{(p+q+r+2)!}=\int_0^1 \!\! ds_1 \int_0^{1-s_1} \!\! ds_2 \ s_1^ps_2^q(1-s_1-s_2)^r.
\end{equation*}
After all a symmetrized trace can be reduced to an ordinary trace with parameter integrals to symmetrize. 

Next we expand $e^{ikX}$ with respect to $\delta X^i$. We can expand $e^{A+B}$ in $B$ as
\begin{equation}
\begin{split}
   e^{A+B}
   &=e^A+\sum_{n=0}^\infty \frac{1}{(n+1)!}\sum_{i=0}^n A^i B A^{n-i} +\sum_{n=0}^\infty \frac{1}{(n+2)!}A^iBA^jA^{n-i-j}+\cdots \\
   &=e^A+\sum_{p,q}\frac{p!q!}{(p+q+1)!} \frac{A^p}{p!} B \frac{A^q}{q!} +\sum_{p,q,r}\frac{p!q!r!}{(p+q+r+2)!} \frac{A^p}{p!} B \frac{A^q}{q!} B \frac{A^r}{r!} + \cdots \\
   &=e^A+\int_0^1 \!\! d\tau \ e^{\tau A} B e^{(1-\tau)A}+\int_0^1 \!\! d\tau_1 \int_0^{1-\tau_1} \!\! d\tau_2 \ e^{\tau_1 A} B e^{\tau_2 A} B e^{(1-\tau_1-\tau_2)A} +\cdots.
\end{split}\label{eq:b3}
\end{equation}
Here parameter integrals appear again. Taking the trace of this equation we have
\begin{equation}
   \mathrm{tr}e^{A+B}=\mathrm{tr}e^A+\mathrm{tr}e^A B +\frac{1}{2}\mathrm{tr}\int_0^1 \!\! d\tau \ e^{\tau A }B e^{(1-\tau)A}B + \cdots. \label{eq:b4}
\end{equation}
Here we use a relation
\begin{equation*}
  \int_0^1\!\!d\tau_1\int_0^{1-\tau_1}\!\!d\tau_2\ \mathrm{tr} e^{(1-\tau_2) A}Be^{\tau_2 A}B
=\int_0^1\!\!d\tau'_1\int_{1-\tau'_1}^1\!\!d\tau'_2\ \mathrm{tr} e^{\tau'_2 A}Be^{(1-\tau'_2) A}B
\end{equation*}
with $\tau'_1=1-\tau_1,\tau'_2=1-\tau_2$. By using these relations we can expand $e^{ikX}$ in a fluctuation $\delta X$. 

Furthermore we need to know matrix components of $e^{Uik\hat{x}U^\dagger}$ in number operator representation in order to calculate charge density formulas for the unstable soliton $X_0^i=Ua^\dagger U^\dagger$ with small fluctuations. $\langle m |e^{ik\hat{x}}| n \rangle $ can be calculated by performing an integral in the equation
\begin{equation*}
\langle m | e^{ik\hat{x}} |n \rangle =\frac{e^{ -\frac{\theta}{4}k^2 }}{\pi\sqrt{m!n!}}\int \!\! d\lambda_1 d\lambda_2 \  (\lambda_1+i\lambda_2)^m(\lambda_1-i\lambda_2)^n e^{-\left(\lambda_1-i\sqrt{\frac{\theta}{2}}k_1\right)^2-\left(\lambda_2-i\sqrt{\frac{\theta}{2}}k_2\right)^2}.
\end{equation*}

Since $UU^\dagger$ in charge density formulas is infinite size, it should be calculated algebraically by using relations
\begin{equation}
   e^{Uik_1\hat{x}U^\dagger} UU^\dagger e^{Uik_2\hat{x}U^\dagger}= Ue^{i(k_1+k_2)\hat{x}}U^\dagger. \label{eq:b6}
\end{equation}
and
\begin{equation}
\begin{split}
&e^{iskX}UU^\dagger e^{i(1-s)kX}\\
&=Ue^{ik\hat{x}}U^\dagger\\
&+\int_0^1 \!\! d\tau \ e^{Ui\tau sk\hat{x}U^\dagger}isk\delta X Ue^{i((1-\tau)s+(1-s))k\hat{x}}U^\dagger \\
&+\int_0^1 \!\! d\tau \ Ue^{i(s+\tau(1-s))k\hat{x}}U^\dagger i(1-s)k\delta X e^{Ui(1-\tau)(1-s)k\hat{x}U^\dagger} \\
&+\int_0^1 \!\! d\tau_1 \ \int_0^1 \!\! d\tau' \ e^{Ui\tau sk\hat{x}U^\dagger}isk\delta X Ue^{i((1-\tau)s+\tau'(1-s))k\hat{x}}U^\dagger i(1-s)k\delta X e^{Ui(1-\tau')(1-s)k\hat{x}U^\dagger} \\
&+\int_0^1 \!\! d\tau_1 \ \int_0^{1-\tau_1} d\tau_2 \ e^{Ui\tau_1 sk\hat{x} U^\dagger}isk\delta X e^{Ui\tau_2 sk\hat{x}U^\dagger}isk\delta X Ue^{i((1-\tau_1-\tau_2)s+(1-s))k\hat{x}}U^\dagger \\
&+\int_0^1 \!\! d\tau_1 \ \int_0^{1-\tau_1} d\tau_2 \ U e^{i(s+\tau_1(1-s))k\hat{x}} U^\dagger i(1-s)k\delta X e^{Ui\tau_2 (1-s)k\hat{x}U^\dagger}i(1-s)k\delta X e^{Ui((1-\tau_1-\tau_2)(1-s)k\hat{x}U^\dagger}.
\end{split}\label{eq:b5}
\end{equation}

Now we have prescriptions to take a symmetrized trace, expand $e^{ikX}$ in $\delta X^i$, write down matrix component of $\langle m |e^{ikX_0} | n \rangle$ for given $m,n$ and deal with an infinite size matrix $UU^\dagger$. Put it all together we can calculate charge density formulas.

\section{Details of charge density calculations}\label{append:c}
In this appendix we show detailed calculations of charge density distribution including contributions of mode $d_i$.

\subsection*{D0-charge}
The D0-charge formula 
\begin{equation*}
   \tilde{J}_0(k)=T_{D0}\mathrm{tr}\left[e^{ikX}\right] 
\end{equation*}
is expanded as
\begin{multline*}
T_{D0}\ \mathrm{tr}\left[e^{Uik\hat{x}U^\dagger+ik\delta X}\right]=T_{D0}\left(1+\frac{2\pi}{\theta}\delta^2(k)\right)\\
+ T_{D0}\mathrm{tr}\left[ik\delta X e^{Uik\hat{x}U^\dagger} + \frac{1}{2} \int_0^1 \!\! d\tau \ ik\delta X  e^{Ui\tau k\hat{x}U^\dagger} ik\delta X  e^{Ui(1-\tau)k\hat{x}U^\dagger} + \cdots \right].
\end{multline*}
For the D0-charge formula there is no need to symmetrize because of cyclic property of trace. Substituting matrix components we have
\begin{equation}
\begin{split}
& \tilde{J}_0(k) = T_{D0}\left(1+\frac{2\pi}{\theta}\delta^2(k)\right) \\
&+T_{D0}e^{-\frac{\theta}{4}k^2}\left(-\theta k^2 d_0 + \frac{-4+\theta k^2}{2\sqrt{2}}\theta k^2 d_1 -\frac{\sqrt{3}}{24}(24-12\theta k^2+\theta^2 k^4)d_2 + \cdots \right)\\
&+T_{D0} \left[ \frac{1}{2}\int_0^1 \!\! d\tau \ e^{-\frac{\theta}{4}k^2\tau^2}\left(-\theta k^2 \right)t_0^2 +\frac{1}{2}\int_0^1 \!\! d\tau \ e^{-\frac{\theta}{4}k^2(\tau^2+(1-\tau)^2)} \left\{ \left(-\theta k^2+\frac{\theta}{4}k^4\right)d_0^2 \right.\right. \\
& +\left(-\theta k^2 +\frac{\theta^2}{4}k^4(2\tau^2-2\tau+3)-\frac{\theta^3}{16}k^6(1+2\tau^2-4\tau^3+2\tau^4)+\frac{\theta^4}{32}k^8(\tau^2-2\tau^3+\tau^4)\right)d_1^2 \\
& + \left.\left. \left(\frac{\sqrt{2}}{4}\theta k^2(1+2\tau-2\tau^2)-\frac{\sqrt{2}}{8}\theta^2k^4(\tau-\tau^2)\right)d_0d_1 + \cdots \right\} \right]\\
&+\cdots .
\end{split}\label{eq:c1}
\end{equation}
The Fourier transform of this gives D0-charge distribution as 
\begin{equation}
\begin{split}
   J_0(x)
&=T_{D0}\left[\delta^2(x)+\frac{1}{2\pi\theta}\right]\\
&+T_{D0}\frac{4}{\pi\theta}\left[e^{-r^2}\left\{(r^2-1)d_0+(r^4-3r^2+1)d_1+\frac{1}{\sqrt{3}}(2r^6-12x^4+15r^2-3)d_2\right\}+\cdots\right] \\
&+T_{D0}\frac{2}{\pi\theta}\left[\int_0^1 \!\! d\tau \ e^{-\frac{r^2}{(1-\tau)^2}}\frac{r^2-(1-\tau)^2}{(1-\tau)^6}t_0^2 \right. \\
&\left. \int_0^1 \!\! d\tau \ e^{-\frac{r^2}{\tau^2+(1-\tau)^2}}\frac{r^4+(-3+4\tau-8\tau^3+4\tau^4)r^2-(-1-2\tau+2\tau^2)(1-2\tau+2\tau^2)}{(1-2\tau+2\tau^2)^5}d_0^2 +\cdots \right]\\
&+\cdots .
\end{split} \label{eq:c2}
\end{equation}

\subsection*{D2-charge}
The D2-charge formula
\begin{equation*}
   \tilde{J}_2(k)=T_{D0}\mathrm{tr}\left[-i[X^1,X^2]e^{ikX}\right]
\end{equation*}
 is expanded as
\begin{equation*}
\begin{split}
-i&\mathrm{tr} [X^1,X^2]e^{ikX} \\
=& \theta \mathrm{tr}UU^\dagger e^{Uik\hat{x}U^\dagger} +\theta \mathrm{tr} \int_0^1 \!\! d\tau\ UU^\dagger e^{Ui\tau k\hat{x} U^\dagger} ik\delta X e^{Ui(1-\tau)k\hat{x}U^\dagger} \\
&+\theta \mathrm{tr}\int_0^1 \!\! d\tau_1 \int_0^{1-\tau_1} \!\! d\tau_2 \ UU^\dagger e^{Ui\tau_1k\hat{x}U^\dagger}ik\delta X e^{Ui\tau_2k\hat{x}U^\dagger} ik\delta X e^{Ui(1-\tau_1-\tau_2)k\hat{x}U^\dagger} \\
&-i\mathrm{tr}\delta[X^1,X^2]e^{Uik\hat{x}U^\dagger}-i\mathrm{tr}\int_0^1 \!\! d\tau \ \delta[X^1,X^2]e^{Ui\tau k\hat{x}U^\dagger} ik\delta X e^{Ui(1-\tau)k\hat{x} U^\dagger} \\
= &2\pi\delta^2(k)+\theta \mathrm{tr}Ue^{ik\hat{x}}U^\dagger ik\delta X \\
&+\frac{1}{2}\theta \mathrm{tr}\int_0^1\!\! d\tau\ Ue^{i(1-\tau)k\hat{x}}U^\dagger ik\delta X U e^{i\tau k\hat{x}}U^\dagger ik\delta X \\
&+\theta \int_0^1\!\! d\tau_1 \int _0^1 \!\! d\tau_2\ \langle 0 | ik\delta X U e^{i(1-\tau_2)k\hat{x}} U^\dagger  ik\delta X | 0 \rangle \\
&-i\mathrm{tr}\delta[X^1,X^2]e^{Uik\hat{x}U^\dagger}-i\mathrm{tr}\int_0^1 \!\! d\tau \ \delta[X^1,X^2]e^{Ui\tau k\hat{x}U^\dagger} ik\delta X e^{Ui(1-\tau)k\hat{x} U^\dagger}.
\end{split} 
\end{equation*}
After component calculations we have
\begin{equation}
\begin{split}
& \tilde{J}_2(k)=\frac{T_{D0}\theta}{2\pi\alpha'}\left[\frac{2\pi}{\theta}\delta^2(k) \right. \\
& +\left\{(1-e^{-\frac{\theta}{4}k^2})-\int_0^1 \!\! d\tau_1 \int_0^{1-\tau_1}\!\! d\tau_2 \ e^{-\frac{\theta}{4}(1-\tau_2)^2k^2}\frac{\theta}{2}k^2 \right\} t_0^2 \\
& +\left\{e^{-\frac{\theta}{4}k^2}\frac{\theta}{2}k^2-\int_0^1 \!\! d\tau \ e^{-\frac{\theta}{4}(\tau^2+(1-\tau)^2)}\frac{\theta}{8}k^2 \left(4-\theta k^2 (1-2\tau)^2 \right)\right\}d_0^2 \\
& +\left\{e^{-\frac{\theta}{8}k^2}\frac{\theta}{2}k^2\left(4-\theta k^2 \right) -\int_0^1 \!\! d\tau \ e^{-\frac{\theta}{4}(\tau^2+(1-\tau)^2)}\frac{\theta}{2}k^2 \left( 1-\frac{\theta}{4}k^2(3-10\tau+10\tau^2) \right.\right. \\
&\phantom{+\{ } \left.\left. +\frac{1}{16}\theta^2k^4(1-8\tau+26\tau^2-36\tau^3+18\tau^4)-\frac{1}{32}\theta^3k^6\tau^2(1-3\tau+2\tau^2)^2\right) \right\}d_1^2 \\
& \left. -\left\{ \int_0^1\!\!d\tau\ e^{-\frac{\theta}{4}(\tau^2+(1-\tau)^2)}\frac{\theta^2k^4}{4\sqrt{2}}\left((1-6\tau+6\tau^2)+\frac{\theta k^2}{2}(\tau-5\tau^2+8\tau^3-4\tau^4)\right)\right\}d_0d_1 +\cdots \right].
\end{split}\label{eq:c3}
\end{equation}
We can see that all of these terms vanish by using equations below.
\begin{equation}
\begin{split}
&\int_0^1 \!\! d\tau_1 \int_0^{1-\tau_1}\!\! d\tau_2 \ e^{-\frac{\theta}{4}(1-\tau_2)^2k^2}\frac{\theta}{2}k^2 = (1-e^{-\frac{\theta}{4}k^2}) \\
&\int_0^1 \!\! d\tau \ e^{-\frac{\theta}{4}(\tau^2+(1-\tau)^2)}\frac{\theta}{8}k^2 \left(4-\theta k^2 (1-2\tau)^2 \right) = e^{-\frac{\theta}{4}k^2}\frac{\theta}{2}k^2 \\
&\int_0^1 \!\! d\tau \ e^{-\frac{\theta}{4}(\tau^2+(1-\tau)^2)}\frac{\theta}{2}k^2 \left( 1-\frac{\theta}{4}k^2(3-10\tau+10\tau^2) \right.\\
&\left. +\frac{1}{16}\theta^2k^4(1-8\tau+26\tau^2-36\tau^3+18\tau^4)-\frac{1}{32}\theta^3k^6\tau^2(1-3\tau+2\tau^2)^2\right) =e^{-\frac{\theta}{4}k^2}\frac{\theta}{8}k^2\left(4-\theta ^2 \right)\\
&\int_0^1\!\!d\tau\ e^{-\frac{\theta}{4}(\tau^2+(1-\tau)^2)}\frac{\theta^2k^4}{4\sqrt{2}}\left((1-6\tau+6\tau^2)+\frac{\theta k^2}{2}(\tau-5\tau^2+8\tau^3-4\tau^4)\right)=0 \\
& \qquad \vdots .
\end{split}\label{eq:c4}
\end{equation}

\subsection*{F-charge}
F-charge formulas can be expanded as
\begin{equation*}
\begin{split}
&\tilde{I}^1(k)
= \frac{T_{D0}g}{2\pi\alpha'}\mathrm{Str}i[X^1,X^2]\dot{X}^2  \\
=& \frac{T_{D0}g}{2\pi\alpha'} i \mathrm{tr}\int_0^1 \!\! ds\ [X^1,X^2] e^{i(1-s)kX} \dot{X}^2 e^{iskX} \\
=& -\frac{T_{D0}g}{2\pi\alpha'} \theta \mathrm{tr} \int_0^1 \!\! ds\ UU^\dagger e^{i(1-s)kX} \dot{X}^2 e^{iskX} +\frac{T_D0}{2\pi\alpha'} i \mathrm{tr} \int_0^1 \!\! ds\ \delta[X^1,X^2] e^{i(1-s)kX} \dot{X}^2 e^{iskX}\\
=&-\frac{T_{D0}g}{2\pi\alpha'}\theta \int_0^1 \!\! ds\ Ue^{ik\hat{x}}U^\dagger \dot{X}^2\\
&-\frac{T_{D0}g}{2\pi\alpha'}\theta \int_0^1 \!\! ds\ \int_0^1 \!\! d\tau \ e^{Ui\tau sk\hat{x}U^\dagger}isk\delta X Ue^{i((1-\tau)s+(1-s))k\hat{x}}U^\dagger \dot{X}^2 \\
&-\frac{T_{D0}g}{2\pi\alpha'}\theta \int_0^1 \!\! ds\ \int_0^1 \!\! d\tau \ Ue^{i(s+\tau(1-s))k\hat{x}}U^\dagger i(1-s)k\delta X e^{Ui(1-\tau)(1-s)U^\dagger} \dot{X}^2\\
&+\frac{T_{D0}g}{2\pi\alpha'}i\mathrm{tr} \int_0^1 \!\! ds\ \delta[X^1,X^2] Ue^{i(1-s)k\hat{x}}U^\dagger \dot{X}^2 e^{Uisk\hat{x}U^\dagger} +\cdots
\end{split}
\end{equation*}
and similar for $\tilde{I}^2(k)$. Here we use \eqref{eq:b6} to deal with $UU^\dagger$. Substituting matrix components to this equation we have 
\begin{equation}
\begin{split}
&\tilde{I}^1(k)\\
=&\frac{T_{D0}g}{2\pi\alpha'}\left[e^{-\frac{\theta}{4}k^2}(-ik_2)\theta^2(\dot{d}_0-\frac{\sqrt{4}}{2}(-4+\theta k^2)\dot{d}_1+\frac{1}{\sqrt{3}}(24-12\theta k^2+\theta^2 k^4)\dot{d}_2 + \cdots ) \right]\\
&+\frac{T_{D0}g}{2\pi\alpha'}\left[\int_0^1 \!\! ds \int_0^1 \!\! d\tau\ e^{-\frac{\theta}{4}\theta^2(1-s\tau)^2k^2}(-ik_2)s t_0\dot{t}_0 \right. \\
& +\left\{\int_0^1 \!\! ds e^{-\frac{\theta}{4}(s^2+(1-s)^2)}\frac{1}{4}ik_2\theta^3 k^2 (-1+s)s \right.\\
&\phantom{+\{ } \left.\left. + \int_0^1 \!\!\ ds \int_0^1 \!\! d\tau\ e^{-\frac{\theta}{4}(1-2\tau s +2\tau^2 s^2)}\frac{1}{2}ik_2 \theta^2 (-4+\theta k^2) s  \right\}d_0 \dot{d}_0 +\cdots \right] \\
&+(\mbox{terms which vanish after integration over $s,\tau$}) \quad +\cdots.
\end{split}\label{eq:c5}
\end{equation}
In similar way $\tilde{I}^2(k)$ is given by
\begin{equation}
\begin{split}
&\tilde{I}^2(k)\\
=&\frac{T_{D0}g}{2\pi\alpha'}\left[e^{-\frac{\theta}{4}k^2}ik_1\theta^2(\dot{d}_0-\frac{\sqrt{4}}{2}(-4+\theta k^2)\dot{d}_1+\frac{1}{\sqrt{3}}(24-12\theta k^2+\theta^2 k^4)\dot{d}_2  + \cdots )\right]\\
&+\frac{T_{D0}g}{2\pi\alpha'}\left[\int_0^1 \!\! ds \int_0^1 \!\! d\tau\ e^{-\frac{\theta}{4}\theta^2(1-s\tau)^2k^2}ik_1s t_0\dot{t}_0 \right. \\
& +\left\{ \int_0^1 \!\! ds e^{-\frac{\theta}{4}(s^2+(1-s)^2)}\frac{1}{4}(-ik_1)\theta^3 k^2 (-1+s)s \right. \\
& \phantom{+\{ } \left.\left.  +\int_0^1 \!\!\ ds \int_0^1 \!\! d\tau\ e^{-\frac{\theta}{4}(1-2\tau s +2\tau^2 s^2)}\frac{1}{2}(-ik_1) \theta^2 (-4+\theta k^2) s  \right\}d_0\dot{d}_0 +\cdots  \right] \\ 
&+(\mbox{terms which vanish after integration over $s,\tau$})\quad +\cdots .
\end{split}\label{eq:c6}
\end{equation}
Here some terms cancel out each other by transforming integral variables as $s \to 1-s$ and $\tau \to 1-\tau$. We can see that the disappearance of total F-charge $\tilde{I}^i(k=0)=0$ is satisfied. F-charge distribution on the D2-brane is given by 
\begin{equation}
\begin{split}
&{I}^1(x)
=\frac{T_{D0}g}{2\pi\alpha'}\left[e^{-r^2}\frac{\pi}{2} x_2 (\dot{d}_0+\sqrt{2}(-1+r^2)\dot{d}_1+\frac{1}{\sqrt{3}}(3-6r^2+2r^4))\dot{d}_2+\cdots \right] \\
&+\frac{T_{D0}g}{2\pi\alpha'}\left[\int_0^1 \!\!ds \int_0^1 \!\! d\tau\ e^{-\frac{r^2}{(-1+s\tau)^2}}\frac{2}{\pi}\frac{sx_2}{(-1+s\tau)^4} t_0\dot{t}_0 \right. \\
& + \left\{\int_0^1 \!\! ds\ e^{-\frac{r^2}{(s^2+(1-s)^2)}} 16\pi x_2 (-1+s)s \frac{r^2-2(1-2s+2s^2)}{(1-s)^2+s^2} \right. \\
& \phantom{+\{ } \left.\left. +\int_0^1 \!\! ds \int_0^1 \!\! d\tau\ e^{-\frac{r^2}{(1-s\tau)^2+s^2\tau^2}}16\pi x_2 s \frac{r^2+(-1+4s^2\tau^2-8s^3\tau^3+4s^4\tau^4)}{(1-s\tau)^2+s^2\tau^2} \right\}d_0\dot{d}_0 +\cdots \right] \\
&+\cdots
\end{split}\label{eq:c7}
\end{equation}
and
\begin{equation}
\begin{split}
&{I}^2(x)
=\frac{T_{D0}g}{2\pi\alpha'}\left[e^{-r^2}\frac{\pi}{2} (-x_1) (\dot{d}_0+\sqrt{2}(-1+r^2)\dot{d}_1+\frac{1}{\sqrt{3}}(3-6r^2+2r^4))\dot{d}_2+\cdots \right] \\
&+\frac{T_{D0}g}{2\pi\alpha'}\left[\int_0^1 \!\!ds \int_0^1 \!\! d\tau\ e^{-\frac{r^2}{(-1+s\tau)^2}}\frac{2}{\pi}\frac{s(-x_1)}{(-1+s\tau)^4} t_0\dot{t}_0 \right. \\
& + \left\{\int_0^1 \!\! ds\ e^{-\frac{r^2}{(s^2+(1-s)^2)}} 16\pi (-x_1) (-1+s)s \frac{r^2-2(1-2s+2s^2)}{(1-s)^2+s^2} \right. \\
&\phantom{+\{ } \left.\left. +\int_0^1 \!\! ds \int_0^1 \!\! d\tau\ e^{-\frac{r^2}{(1-s\tau)^2+s^2\tau^2}}16\pi (-x_1) s \frac{r^2+(-1+4s^2\tau^2-8s^3\tau^3+4s^4\tau^4)}{(1-s\tau)^2+s^2\tau^2} \right\}d_0\dot{d}_0+\cdots \right] \\
&+\cdots.
\end{split}\label{eq:c8}
\end{equation}
We can see that $\vec{I}(x)$ takes the form of $f(r^2)\begin{pmatrix}x_2 \\ -x_1 \end{pmatrix}$.

\subsection*{Energy density}
In this appendix we calculate energy density distribution. First we divide the energy density formula into four parts as
\begin{align*}
  \tilde{T}_1(k)&=T_{D0}\frac{g}{2}\mathrm{Str}((\dot{X}^1)^2+(\dot{X}^2)^2)e^{ikX} \\
  \tilde{T}_2(k)&=T_{D0}\mathrm{tr}e^{ikX} \\
  \tilde{T}_3(k)&=T_{D0}\frac{ig\theta}{(2\pi\alpha')^2}\mathrm{tr}[X^1,X^2]e^{ikX} \\
  \tilde{T}_4(k)&=T_{D0}\frac{g^2}{2(2\pi\alpha')^2}\mathrm{Str}[X^1,X^2]^2e^{ikX}. 
\end{align*}
Turning on fluctuations around the unstable soliton $\tilde{T}_1(k),\tilde{T}_2(k),\tilde{T}(k)_3$ can be expanded as
\begin{equation*}
\begin{split}
&\tilde{T}_1(k)
=T_{D0}\frac{g}{2}\int_0^1\!\!ds\ \mathrm{tr} \dot{X}^i e^{Uisk\hat{x}U^\dagger} \dot{X}^i e^{Ui(1-s)k\hat{x}U^\dagger} \\
&+T_{D0}\frac{g}{2}\int_0^1\!\!ds \int_0^1\!\!d\tau\ \dot{X}^i e^{Uisk\hat{x}U^\dagger} \dot{X}^i e^{Ui(1-s)k\hat{x}U^\dagger} i(1-s)k\delta X e^{Ui(1-s)(1-\tau)k\hat{x}U^\dagger}\\
&+T_{D0}\frac{g}{2}\int_0^1\!\!ds \int_0^1\!\!d\tau\ \dot{X}^i e^{Uis\tau k\hat{x}U^\dagger} isk\delta X e^{Uis(1-\tau)k\hat{x}U^\dagger} \dot{X}^i e^{Ui(1-s)k\hat{x}U^\dagger} +\cdots
\end{split}
\end{equation*}
and
\begin{equation*}
  \tilde{T}_2(k)=T_{D0}\left[e^{Uik\hat{x}U^\dagger}+\mathrm{tr}ik\delta X e^{Uik\hat{x}U^\dagger} +\frac{1}{2}\int_0^1\!\!d\tau\ \mathrm{tr}ik\delta X e^{Ui\tau k\hat{x}U^\dagger}ik\delta X e^{Ui(1-\tau)k\hat{x}U^\dagger} +\cdots \right]
\end{equation*}
and
\begin{equation*}
\begin{split}
&\tilde{T}_3(k)
=T_{D0}\frac{ig\theta}{(2\pi\alpha')^2}\left[-\theta \mathrm{tr}UU^\dagger e^{ikX} +i\mathrm{tr}\delta[X^1,X^2]e^{ikX} \right] \\
&=T_{D0}\frac{ig\theta}{(2\pi\alpha')^2}\left[-\theta\mathrm{tr}Ue^{ik\hat{x}}U^\dagger -\theta \mathrm{tr}Ue^{ik\hat{x}}U^\dagger ik\delta X \right.\\
&\phantom{T_{D0}}-\theta\mathrm{tr}\int_0^1\!\!d\tau\ Ue^{i(1-\tau)k\hat{x}}U^\dagger ik\delta X Ue^{i\tau k\hat{x}}U^\dagger ik\delta X 
- \int_0^1\!\!d\tau_1\int_0^{1-\tau_1}\!\!d\tau_2\ \langle 0 |ik\delta X Ue^{i(1-\tau_2)k\hat{x}}U^\dagger ik\delta X | 0 \rangle \\
&\phantom{T_{D0}}\left.  +i\mathrm{tr}\delta[X^1,X^2]e^{Uik\delta xU^\dagger}+i\mathrm{tr}\int_0^1\!\!d\tau\ e^{Ui\tau k\hat{x} U^\dagger}e^{Ui(1-\tau)k\hat{x}U^\dagger} +\cdots \right].
\end{split} 
\end{equation*}
Before expanding $\tilde{T}_4(k)$ in fluctuations we perform some calculations which is needed later. Operating $UU^\dagger$ from the left to \eqref{eq:b5} we have
\begin{equation*}
\begin{split}
&\int_0^1\!\!ds\ \mathrm{tr}UU^\dagger e^{iskX}UU^\dagger e^{i(1-s)kX}
=\mathrm{tr}e^{ik\hat{x}}+\mathrm{tr}ik\delta X Ue^{ik\hat{x}}U^\dagger \\
&+\int_0^1\!\!d\tau \int_0^1\!\!d\tau' \int_0^1\!\!ds\ \mathrm{tr}isk\delta X Ue^{i((1-\tau)s+\tau'(1-s))k\hat{x}}U^\dagger i(1-s)k\delta X Ue^{i((1-\tau')(1-s)+\tau s)k\hat{x}}U^\dagger \\
&+2\int_0^1\!\!d\tau_1\int_0^{1-\tau_1}\!\!d\tau_2\int_0^1\!\!ds\ \mathrm{tr}isk\delta X e^{Ui\tau_2 k\hat{x}}U^\dagger isk\delta X Ue^{i(1-\tau_2 s)k\hat{x}}U^\dagger +\cdots.
\end{split}
\end{equation*}
By using this equation we can expand $\tilde{T}_4(k)$ as
\begin{equation*}
\begin{split}
&\tilde{T}_4(k)
=-T_{D0}\frac{g^2}{2(2\pi\alpha')^2}\theta^2 \int_0^1 \!\! ds\ \mathrm{tr}UU^\dagger e^{iskX} UU^\dagger e^{i(1-s)kX} \\
&+T_{D0}\frac{g^2}{2(2\pi\alpha')^2}i\theta \int_0^1 \!\!\ ds \mathrm{tr} UU^\dagger e^{iskX} \delta[X^1,X^2] e^{i(1-s)kX}+i\theta \int_0^1 \!\! ds\ \mathrm{tr}\delta[X^1,X^2]e^{iskX}UU^\dagger e^{i(1-s)kX} \\
&+T_{D0}\frac{g^2}{2(2\pi\alpha')^2}\int_0^1 \!\! ds\ \mathrm{tr}\delta[X^1,X^2]e^{iskX} \delta[X^1,X^2] e^{i(1-s)kX} \\
=&-T_{D0}\frac{g^2}{2(2\pi\alpha')^2}\theta^2\left[\mathrm{tr}e^{ik\hat{x}}+\mathrm{tr}ik\delta X Ue^{ik\hat{x}}U^\dagger \right. \\
&\phantom{-T_{D0}-\theta^2}+\int_0^1\!\! d\tau \int_0^1\!\! d\tau' \int \!\!ds \ \mathrm{tr}isk\delta X Ue^{i(1-\tau s+\tau'(1-s))k\hat{x}}U^\dagger i(1-s)k\delta X Ue^{i((1-\tau')(1-s)+\tau s)}U^\dagger \\
&\phantom{-T_{D0}-\theta^2}\left. +2\int_0^1\!\! d\tau_1 \int_0^{1-\tau_1}\!\! d\tau_2 \int_0^1\!\! ds\ \mathrm{tr}isk\delta X e^{Ui\tau_2 skU^\dagger}isk\delta X Ue^{i(1-\tau_2s)ks}U^\dagger \right] \\  
&+T_{D0}\frac{g^2}{2(2\pi\alpha')^2}2i\theta\left[\mathrm{tr}\delta [X^1,X^2]Ue^{ik\hat{x}}U^\dagger \right.\\
&\phantom{-T_{D0}2i\theta}+\int_0^1\!\!d\tau\int_0^1\!\!ds\ \mathrm{tr}\delta[X^1,X^2]e^{Ui\tau sk\hat{x}U^\dagger}isk\delta X Ue^{i((1-\tau)s+(1-s))}U^\dagger \\
&\phantom{-T_{D0}2i\theta}\left. +\int_0^1\!\!d\tau\int_0^1\!\!ds\ \mathrm{tr}\delta[X^1,X^2]Ue^{i(s+\tau(1-s))k\hat{x}}U^\dagger i(1-s)k\delta X Ue^{i(1-\tau_2 s)k\hat{x}}U^\dagger \right]\\
&\phantom{-T_{D0}} T_{D0}\frac{g^2}{2(2\pi\alpha')^2} +\int_0^1\!\!ds\ \mathrm{tr} \delta[X^1,X^2]e^{isk\hat{x}}\delta[X^1,X^2]e^{i(1-s)k\hat{x}} +\cdots.
\end{split}
\end{equation*}
Substituting matrix components $\tilde{T}_2+\tilde{T}_3-\tilde{T}_4(k)$ becomes
\begin{equation*}
\begin{split}
&\tilde{T}_2(k)+\tilde{T}_3-\tilde{T}_4(k)
=T_{D0}\left[1+\frac{2\pi}{\theta}\delta^2(k)\left(1-\frac{\theta^2 g^2}{2(2\pi\alpha')^2}\right)\right]\\
&+T_{D0}\left[\left(1-\frac{g^2\theta^2}{2(2\pi\alpha')^2}\right)e^{-\frac{\theta}{4}k^2}\theta k^2\left(-d_0 + \frac{-4+\theta k^2}{2\sqrt{2}} d_1 -\frac{\sqrt{3}}{24}(24-12\theta k^2+\theta^2 k^4)d_2 + \cdots \right) \right] \\
&+T_{D0}\left[\left\{-2\frac{g^2\theta^2}{2(2\pi\alpha')^2}e^{-\frac{\theta}{4}k^2}-\frac{1}{2}\int_0^1\!\!d\tau\ e^{-\frac{\theta}{4}(1-\tau)^2k^2}\theta k^2 \right.\right.\\
&\phantom{T_{D0}\ }\left.-\frac{g^2\theta^2}{2(2\pi\alpha')^2}\int_0^1\!\!d\tau_1 \int_0^{1-\tau_1}\!\!d\tau_2 \int_0^1\!\! ds\ e^{-\frac{\theta}{4}(1-s\tau_2)^2 k^2}s^2\theta k^2  \right\}t_0^2 \\
&\phantom{T_{D0}}+\left\{\frac{1}{8}\left(1-2\frac{\theta^2g^2}{2(2\pi\alpha')^2}\right)\int_0^1\!\!d\tau\ e^{-\frac{\theta}{4}(\tau^2+(1-\tau)^2)k^2}\theta k^2(-4+\theta k^2) \right. \\
&\phantom{T_{D0}\ }-\frac{1}{4}\frac{g^2\theta^2}{2(2\pi\alpha')^2}\int_0^1\!\!d\tau \int_0^1\!\!d\tau' \int_0^1\!\!ds\ e^{-\frac{\theta}{4} a k^2}s(-1+s)\theta k^2(-4+\theta k^2)  \\
&\phantom{T_{D0}\ }\left.\left.+\frac{1}{2}\frac{g^2\theta^2}{2(2\pi\alpha')^2}\int_0^1\!\!d\tau_1 \int_0^1\!\!d\tau_2 \int_0^1\!\!ds\ e^{-\frac{\theta}{4}(1-2s\tau_2+2s^2\tau_2^2)k^2}s^2\theta k^2 (-4+\theta k^2) \right\}d_0^2 +\cdots \right] +\cdots ,
\end{split}
\end{equation*}
and $\tilde{T}_1(k)$ becomes
\begin{equation*}
\begin{split}
&\tilde{T}_1(k)=
T_{D0}\frac{g}{2}\left[\int_0^1\!\!ds\ e^{-\frac{\theta}{4}(1-s)^2k^2}\theta \dot{t}_0^2-\frac{1}{2}\int_0^1\!\!ds\ e^{-\frac{\theta}{4}(s^2+(1-s)^2)k^2}(-4+(s^2+(1-s)^2)\theta k^2)\theta \dot{d}_0^2 +\cdots \right] \\
&+T_{D0}\frac{g}{2}\left[\int_0^1\!\!d\tau \int_0^1\!\!ds\ e^{-\frac{\theta}{4}(1+2s+s^2(1+\tau^2))k^2} s(-1+s) \theta k^2 t_0 \theta \dot{t}_0\dot{d}_0 \right.\\
&\phantom{T_{D0}}\left. +\frac{1}{2\sqrt{2}}\int_0^1\!\!d\tau \int_0^1\!\!ds\ e^{-\frac{\theta}{4}(1-2s+s^2(1+\tau)^2)k^2}(1-s)^2s^2\tau \theta^2 k^4 t_0 \theta \dot{t}_0\dot{d}_1 +\cdots  \right] + \cdots .
\end{split}
\end{equation*}
where $a=((1-\tau)s+\tau'(1-s))^2+((1-\tau')(1-s)+\tau s)^2$. The Fourier transform of this gives energy density $T_2(x)+T_3(x)-T_4(x)$ as
\begin{equation*}
T_2(x)+T_3(x)-T_4(x)|_\mathrm{0th}
=T_{D0}\left[\delta^2(x)+\frac{1}{2\pi\theta}\left(1-\frac{\theta^2 g^2}{2(2\pi\alpha')^2}\right)\right]
\end{equation*}
\begin{equation*}
T_2(x)+T_3(x)-T_4(x)|_\mathrm{1st}
=T_{D0}\left[\frac{4}{\pi\theta}\left(1-\frac{g^2\theta^2}{2(2\pi\alpha')^2}\right)e^{-r^2}\left((r^2-1)d_0+\sqrt{2}(r^4-3r^2+1)d_1+ \cdots \right) \right]
\end{equation*}
\begin{equation}
\begin{split}
&T_2(x)+T_3(x)-T_4(x)|_{t_0^2}
=T_{D0}\left\{-\frac{2}{\pi\theta}\frac{g^2\theta^2}{2(2\pi\alpha')^2}e^{-r^2}+\int_0^1\!\!d\tau\ e^{-\frac{r^2}{(1-\tau)^2}}\frac{2(r^2-(1-\tau)^2)}{\pi\theta(1-\tau)^6} \right.\\
&\phantom{T_{D0}}\left.+\frac{g^2\theta^2}{2(2\pi\alpha')^2}\int_0^1\!\!d\tau_1 \int_0^{1-\tau_1}\!\!d\tau_2 \int_0^1\!\! ds\ e^{-\frac{r^2}{(1-s\tau_2)^2}}\frac{4s^2(r^2-(1-s\tau_2)^2)}{\pi\theta(1-s\tau_2)^6}\right\}t_0^2
\end{split}\label{eq:c9}
\end{equation}
\begin{equation*}
\begin{split}
&T_2(x)+T_3(x)-T_4(x)|_{d_0^2}\\
=&T_{D0}\left\{\left(1-2\frac{\theta^2g^2}{2(2\pi\alpha')^2}\right)\int_0^1\!\!d\tau\ e^{-\frac{r^2}{(\tau^2+(1-\tau)^2)}} \right.\\
&\phantom{T_{D0}T_{D0}}\frac{2(r^4+r^2(-3+4\tau-8\tau^3+4\tau^4)-(-1-2\tau+2\tau^2)((1-\tau^2)+\tau^2))}{\pi\theta(\tau^2+(1-\tau^2))} \\
&\phantom{T_{D0}}-\frac{\theta^2g^2}{2(2\pi\alpha')^2}\int_0^1\!\!d\tau \int_0^1\!\!d\tau' \int_0^1\!\!ds\ e^{-\frac{r^2}{a}}\frac{4s(1-s)(r^4-4ar^2+a^2(r^2+2)-a^3)}{\pi\theta a^5}  \\
&\phantom{T_{D0}}-\frac{g^2\theta^2}{2(2\pi\alpha')^2}\int_0^1\!\!d\tau_1 \int_0^{1-\tau_1}\!\!d\tau_2 \int_0^1\!\!ds\ e^{-\frac{r^2}{(1-2s\tau_2+2s^2\tau_2^2)}}8s^2 \\
&\left.\phantom{T_{D0}T_{D0}}\frac{-r^4+r^2(3-4s\tau_2+8s^3\tau_2^3-4s^4\tau_2^4)+(-1-2s\tau_2+2s^2\tau_2^2)(1-2s\tau_2+2s^2\tau_2^2)}{\pi\theta(1-2s\tau_2+2s^2\tau_2^2)^5} \right\}d_0^2
\end{split}
\end{equation*}
where $a=(1-2\tau'+2\tau^{'2}+2s^2(-1+\tau+\tau')^2-2s(-1+\tau+\tau')(-1+2\tau'))$. The Fourier transform of $\tilde{T}(k)_1$ gives $T_1(x)$ as
\begin{equation}
\begin{split}
&T_1(x)=
T_{D0}\frac{g}{2}\left[\int_0^1\!\!ds\ e^{-\frac{r^2}{(1-s)^2}}\frac{1}{\pi(1-s)^2} \dot{t}_0^2-e^{-\frac{r^2}{(s^2+(1-s)^2)}}\frac{2r^2}{\pi((s^2+(1-s)^2)^2)}\dot{d}_0^2 +\cdots \right] \\
&+T_{D0}\frac{g}{2}\left[\int_0^1\!\!d\tau \int_0^1\!\!ds\ e^{-\frac{r^2}{1+2s+s^2(1+\tau^2)}} \frac{s(-1+s)(-r^2+(1+2s+s^2(1+\tau^2)))}{\pi(1+2s+s^2(1+\tau^2))^3} t_0 \dot{t}_0\dot{d}_0 \cdots \right] +\cdots.
\end{split}\label{eq:c10}
\end{equation}

\section{T-duality between D0/D2-brane and intersecting D1-strings}\label{append:d}
In this appendix we consider the T-duality between a D0/D2-brane and intersecting D1-strings. Intersecting D1-strings are unstable due to existence of a localized tachyon around the intersecting point and their recombination or reconnection\footnote{Brane recombination have been discussed also in context of string phenomenology, for example in \cite{0203160,0207234}.} is realized by the tachyon condensation. Such a phenomenon called recombination or reconnection corresponds to dissolution of a D0-brane into a D2-brane in a T-dual language. In the Seiberg-Witten limit dissolution can be described by $U(1)$ noncommutative gauge theory, and reconnection can be described by $SU(2)$ Yang-Mills theory.

A D0/D2-brane with a background B-field is T-dual to intersecting D1-strings at a finite angle $\psi$ relating to the strength of the magnetic field $B$ as
\begin{equation}
   \tan \psi = \frac{1}{b}, \quad b=\frac{2\pi\alpha' B}{g}. \label{eq:tdual}
\end{equation}
According to CFT calculations \cite{9910263,0005283} mass spectra of bosonic excitations of 0-2 strings in a D0/D2-brane system and 1-1' strings in a D1-string system are respectively
\begin{equation*}
   m^2_{i(\mathrm{D0/D2})}=\frac{(2i+1)\nu}{2\alpha'} \ i=-1,1,2,\cdots \qquad m^2_{j(\mathrm{D1/D1})}=\frac{(2j+1)\psi}{2\pi\alpha'} \ j=-1,1,2,\cdots
\end{equation*}
except those of order $1/\alpha'$. Here $\nu$ is defined as
\begin{equation*}
   e^{2\pi i\nu}=\frac{-1+ib}{1+ib},
\end{equation*}
and is related to the intersecting angle as
\begin{equation*}
   \psi =\pi\nu.
\end{equation*}
Thus mass spectra in these two systems are identical. In the Seiberg-Witten limit $b$ goes to infinity as $b\to \epsilon^{-1/2}$ and $\nu$ acts like $\frac{1}{\pi b}$. 
Thus the T-dual relation \eqref{eq:tdual} indicates that in the Seiberg-Witten limit the intersecting angle $\psi$ should go to zero as $\psi \to \epsilon^{1/2}$. This is nothing but the parallel limit of intersecting D1-strings. In the parallel limit recombination or reconnection of D1-strings can be described by the 1+1 dimensional $SU(2)$ Yang-Mills theory \cite{0303204,0304237}. We can say that these studies are T-dual versions of our study in this paper.

\EPSFIGURE[ht]{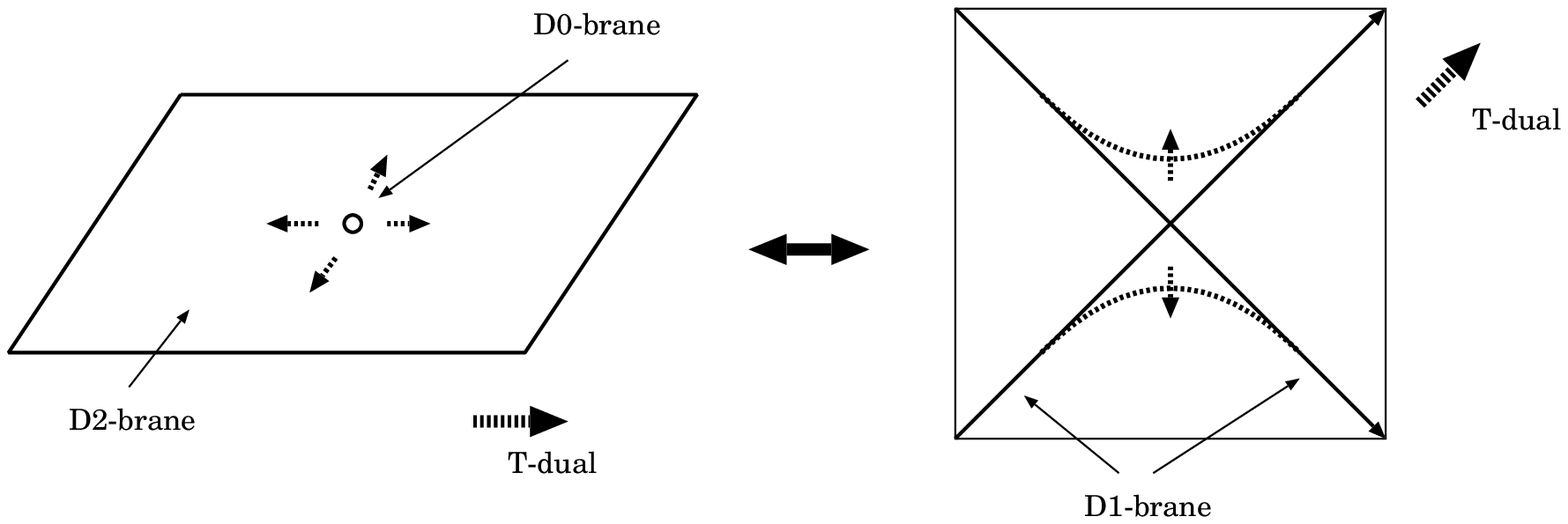,width=\linewidth}{T-duality between a D0/D2-brane and intersecting D1-strings}

\end{document}